\documentclass[prd,aps,floats,twocolumn,nofootinbib]{revtex4-1}
\usepackage{slashed}
\usepackage{mathtools}
\usepackage{amsfonts}
\usepackage{amssymb}
\usepackage{epsfig}
\usepackage{rotating}
\usepackage{url}
\usepackage{times}
\usepackage{color}
\usepackage{bm}
\usepackage{xcolor,colortbl}
\usepackage{hyperref}
\usepackage[alsoload=hep]{siunitx}
\usepackage{enumitem}
\usepackage{fancyhdr}
\usepackage{anyfontsize}
\usepackage{wasysym}
\usepackage{empheq}

\newcommand{\nn}{\nonumber}

\newcommand{\be}{\begin{equation}}
\newcommand{\ee}{\end{equation}}
\newcommand{\bea}{\begin{eqnarray}}
\newcommand{\eea}{\end{eqnarray}}

\newcommand{\plk}{\mathfrak{h}}
\newcommand{\fn}{\footnote}

\begin{document}

%%%%%%%%%%%%%%%%%%%%%%%%%%%%%

\title{Connection between cosmological time and the constants of Nature}
%Time, no time and too many times}
\author{Jo\~{a}o Magueijo}
\email{j.magueijo@imperial.ac.uk}
\affiliation{Theoretical Physics Group, The Blackett Laboratory, Imperial College, Prince Consort Rd., London, SW7 2BZ, United Kingdom}

\date{\today}

\begin{abstract}
We examine in greater detail the proposal that time is the conjugate of the constants of nature. 
Fundamentally distinct times are associated with different constants, a situation often found in
``relational time'' settings. We show how in regions dominated by a single constant the Hamiltonian constraint 
can be reframed as a Schrodinger equation in the corresponding time, solved in the connection representation by outgoing-only monochromatic plane waves moving in a ``space'' that generalizes the Chern-Simons functional (valid for the equation of state $w=-1$) for other $w$.
We pay special attention to the issues of unitarity and the measure employed for the inner product. 
Normalizable superpositions can be built, including solitons, ``light-rays'' and coherent/squeezed states saturating a Heisenberg
uncertainty relation between constants and their times. A healthy classical limit is obtained for factorizable coherent states,
both in mono-fluid and multi-fluid situations. For the latter, 
we show how to deal with transition regions, where one is passing on the baton from one time to another, 
and  investigate the fate of the subdominant clock. For this purpose minisuperspace is best seen as a dispersive medium, with packets moving with 
a group speed distinct from the phase speed. We show that the motion of the packets' peaks reproduces the classical limit even
during the transition periods, and for subdominant clocks once the transition is over.
Deviations from the coherent/semi-classical limit are expected in these cases, however. 
The fact that we have recently transitioned from a decelerating to an accelerating Universe renders this proposal potentially testable, 
as explored elsewhere. 
\end{abstract}

\maketitle

\section{Introduction}

The problem of time in General Relativity~\cite{Isham,Kuch,marina,rovelli} and the mystery of the origin and value of the constants of nature~\cite{Barrowtip}
are two well-known connundrums  embedded in the foundations of physics. In~\cite{Myletter}
we suggested that  they could be inextricably intertwined. 
A priori this should certainly be the case. 
Physical time concerns relational {\it change}.
In contrast, the constants of nature, 
if true to their name, are the hallmarks of {\it immutability}.  
We could therefore expect that time and the constants are conjugate dynamical variables,
or at least complementary in a quantum sense.
Should we, therefore, promote the constants to observables, with their complementaries providing physical definitions
of time?

Naturally, 
%the devil is in the detail, and
a great many questions pour in. Foremost, given the plethora of constants (some more fundamental than others~\cite{Duff}), 
we have to settle on whether we should contend  with different physical times, or if, instead, a select constant is the progenitor of a single time. In~\cite{Myletter} we suggested following the first route. A clock is built with what is at hand.  Depending on the constant(s) dominating the dynamics, different phase space regions should employ different times. Thus, we are led to times variously conjugate to the cosmological constant, $\Lambda$,  the gravitational constant, $G_N$, or even the speed of light, $c$. Within such a pragmatic approach  we  need to know how to pass on the baton from one clock to another. This adjustment of clocks should be seen as a physical  feature of our world.

Our proposal may sound radically new, but in fact it is rooted in well-known literature. 
In the context of $\Lambda$ it follows directly from findings in unimodular gravity~\cite{unimod1} (in particular in the 
formulation of 
Henneaux and Teitleboim~\cite{unimod}), 
where the conjugate of $\Lambda$ is identified with a time variable (unimodular time~\cite{unimod,Bombelli,UnimodLee2},
 a 4D version of Misner's volume time~\cite{misner}). 
The proposal in~\cite{Myletter} amounts to extending these ideas and applying them to constants other than $\Lambda$.
A further point of contact is the concept of Chern-Simons time~\cite{Chopin}, related to York time~\cite{york}, to be reinterpreted here. 

%The implications will be investigated further here. 

The plan of this paper is as follows.
In Section~\ref{beyondMSS} we start by reviewing the origins of our proposal:  the fully diffeomorphism invariant formulation of unimodular  gravity~\cite{unimod1,unimod}, i.e. the formulation~\cite{unimod} which, contrary to its name, does {\it not} restrict the theory to unimodular diffeomorphisms. Our construction 
can be seen as a generalization of the prescription found in~\cite{unimod}, targeting constants other than $\Lambda$. A reduction to minisuperspace is then presented in Section~\ref{MSSred}, recovering the formalism in~\cite{Myletter}. We quantize the problem and present  general solutions for a single perfect fluid.

In Section~\ref{Lambda} we make further connections with previous literature by illustrating this procedure in the case of pure $\Lambda$. We show that in the connection representation 
the monochromatic plane waves move in a time (proportional to the time defined in~\cite{unimod}) conjugate to $1/\Lambda$ (seen as a ``frequency''), and in a space which is
the Chern-Simons (CS) functional (so that the spatial part of the wave is the real CS state~\cite{kodama}). 
%it  leads to the real Chern-Simons (CS) state. However this now can be seen as a monochromatic plane wave in a setting where time is the conjugate to $1/\Lambda$ (the ``frequency''), and space is what Smolin and Soo called Chern-Simons time. Given that the waves move at the speed of light, the two approaches can be confused in most cases.  
But crucially,  we can now superpose the monochromatic plane waves into
normalizable solutions, as we show in Section~\ref{genstats}, where we identify solitons, light rays and coherent/squeezed states. 
%However the waves are not plane, nor move at a constant speed in terms of the connection
%variable. The Chern-Simons time/space is like a linearizing variable in DSR.)
%\item 
In Section~\ref{Genfluids} we present a straightforward generalization to Universes dominated by 
radiation and fluids with generic equation of state.  In each of these the constant of choice is different, so that the chosen time is
different. 
Having a quantum time variable and the ability to find peaked wave-packets is instrumental in finding the correct classical limit.
We explain how this is achieved by coherent states in Section~\ref{classlimit}, once classical cosmology in rephrased in
the connection representation and with constant-times. 

The rest of the paper is spent formalizing  how to change the clock  in multi-fluid situations, 
as already sketched in~\cite{Myletter} and reviewed in Section~\ref{multitime}. 
In Section~\ref{MSSDSR} we show how minisuperspace can be seen as a dispersive medium, 
with wave-numbers that can be position and frequency dependent. By examining the associated
group speed we can then find the equations of motion
of the peak of suitable wave functions. Using this technique (and assuming that the wave function remains peaked) we prove 
in Section~\ref{crossover}
 that the correct semi-classical limit is still obtained in cross-over regions. We also illuminate the fate of the 
minority clock once the handover of clocks is completed. In Section~\ref{why-swap} we formalize the reasons why a clock should indeed
be built with ``what is at hand'', exposing the limitations of minority clocks.

In a concluding Section 
%we briefly explain how our construction can be extended beyond minisuperspace, and
we summarize our findings and discuss their ultimate implications.

\section{Deconstantization}\label{beyondMSS}

The construction in this paper can be seen as a generalization of the fully diffeomorphism invariant formulation of unimodular  gravity~\cite{unimod1,unimod} (i.e. the formulation which does {\it not} restrict to unimodular diffeomorphisms). 
The unimodular theory of gravity is renown for converting the cosmological constant from a fixed parameter in the 
Lagrangian into an integration constant, owing its constancy to an equation of motion, i.e. demoting it to 
an on-shell-constant only. To this process we will call ``deconstantization''.
% and it reveals a time variable. 

In the formulation of~\cite{unimod}, which we now review, one adds to a ``base theory''\footnote{Which here will be just the standard action 
of General Relativity plus a cosmological constant or whatever matter content; but this could be applied to any other ``base theory''.}
% we consider, but could be any other ``base theory'') 
with action $S_0$ a new term:
\be\label{Utrick}
S_0\rightarrow S=S_0- \int d^4 x \Lambda \partial_\mu T^\mu_U=
S_0+ \int d^4 x (\partial_\mu\Lambda)  T^\mu_U
\ee
(the two  equivalent up to a boundary term).
$\Lambda$ is a scalar and $T_U^\mu$ is a vector {\it density}, so that the 
added term is indeed diffeomorphism invariant (note that for a density $\nabla_\mu T^\mu_U =\partial_\mu T^\mu_U$,
and that the integrand has the correct weight for the integral to be a scalar). Upon a 3+1 split,  $T_U^0$ becomes the canonical conjugate of $\Lambda$, and it was pointed out in~\cite{unimod,Bombelli,UnimodLee2} that $T_U^0$ can be used to craft a definition
of time. The density $T_U^\mu$ does not appear in $S_0$,
therefore:
\be\label{ConstL}
\frac{\delta S}{\delta T_U ^\mu}=0\implies \partial_\mu \Lambda =-\frac{\delta S_0}{\delta T^\mu_U}=0,
\ee
i.e. the promised  on-shell constancy of $\Lambda$. 
The other equation of motion is:
\be\label{eom2}
\frac{\delta S}{\delta \Lambda }=0\implies \partial_\mu  T^\mu_U=\frac{\delta S_0}{\delta \Lambda}=-\frac{\sqrt{-g}}{8\pi G_N}.
\ee
As suggested in~\cite{unimod}, 
the  gauge invariance $T^\mu\rightarrow T^\mu + \epsilon^\mu$ (with 
$\partial_\mu\epsilon^\mu=0$)  implies that we should only consider as physical the zero-mode
of $T^0$, and none of its other components (this point is irrelevant in a minisuperspace reduction). Given (\ref{eom2}) we find that on-shell
this is nothing but unimodular time~\cite{unimod,Bombelli,UnimodLee2} (a 4D version of Misner's volume time~\cite{misner}).  The second equation of motion, Eq.~ (\ref{eom2}),
should then be seen as the 
``time formula'' of the theory.

The same prescription could be applied to any other
supposed constant of nature appearing in $S_0$ (leading to the proposal in~\cite{Myletter}, as we will show).
% (with the ``unimodular'' associations excised). 
For a vector of $D$ constants $\bm\alpha$ we should take: 
\be
S_0\rightarrow S=S_0- \int d^4 x\, {\bm \alpha} \cdot \partial_\mu {\bm T}_{\bm \alpha}^\mu
\ee
where the dot denotes the Euclidean inner product in $D$ dimensional space. 
As with $\Lambda$ above, we obtain two extra equations of motion:
\bea
\frac{\delta S}{\delta {\bm T}_{\bm\alpha } ^\mu}=0&\implies& \partial_\mu \bm\alpha=0\\
\frac{\delta S}{\delta \bm\alpha }=0&\implies& \partial_\mu {\bm T}^\mu_{\bm\alpha}=\frac{\delta S_0}{\delta \bm\alpha}.
\eea
These are the on-shell constancy of $\bm \alpha$ 
and generalized  time formulae (several examples of which will be studied in Section~\ref{sec-timef}).
We may either take one constant at a time (with alternative options), or consider a multi-constant setting
with concurrent multiple definitions of time. This is not unusual in relational time formulations (see for example~\cite{Gielen}
for the implications this has for the singularity problem). 
We explain later how such concurrent times become a function of each other classically (Section~\ref{sec-timef}) 
and semi-classically (Section~\ref{crossover}), despite being intrinsically different, off-shell and quantum mechanically. 

%[BUILDING SITE ENTRY POINT]

%The fact that we used $\phi=3/\Lambda$ instead of $\Lambda$ is an expression of the general invariance of this formalism under canonical transformations.
% particularly point transformations. 

Obviously a function of a constant is also a constant, so there is a {\it classical} ambiguity in this construction.
It leads to theories related by a canonical transformation:
\bea
{\bm \alpha}\rightarrow {\bm\beta}(\bm \alpha).
\eea
%knowing that if $\bm \alpha$ are constants on-shell, then so are the  $\bm\beta$.
As with any such canonical transformation, the conjugates transform according to
\bea
 {\bm T}^0_{\bm\beta}=\frac{\delta \bm\alpha}{\delta \bm\beta}{\bm T}^0_{\bm\alpha}.
\eea
More generally, on-shell, we have that:
%The conjugate times' equation transforms according to the inverse Jacobian, since:
%[wait; there must be the usual integrability condition; check but this is a detail]
%since *** transforms according to the 
\be
\partial_\mu {\bm T}^\mu_{\bm\beta}=\frac{\delta S_0}{\delta \bm\beta}=
\frac{\delta \bm\alpha}{\delta \bm\beta}\frac{\delta S_0}{\delta \bm\alpha}=
\frac{\delta \bm\alpha}{\delta \bm\beta}\partial_\mu {\bm T}^\mu_{\bm\alpha}
\ee
so that, using $\partial_\mu{\bm \alpha}=\partial_\mu{\bm\beta}=0$, we have:
% (up to an irrelevant integration constant):
\bea
 {\bm T}^\mu_{\bm\beta}=\frac{\delta \bm\alpha}{\delta \bm\beta}{\bm T}^\mu_{\bm\alpha}.
\eea
All theories generated by an arbitrary choice of $\bm \beta(\bm \alpha)$ are classically (or ``on-shell'') equivalent between themselves (and to GR).
%The first Hamilton equation implies that $\bm \alpha$ (or any function $\bm \beta$ thereof) is a constant. The other field equations then reduce to GR. The second Hamilton equation provides an expression for ${\bm T}^\mu _{\bm \beta}$, and regardless of the choice of $\bm\beta$ they are all a function of each other: classically there is only one time. 
Yet their quantum mechanics is very different. For example, a state can only be coherent and factorizable for one of the choices of $\bm\beta(\bm\alpha)$. The inner product we will propose is also not invariant under such transformations. 
This will be essential in deriving the simplest quantum theory later.
% and in particular in relating the work of~\cite{unimod}

We note that several works found in the previous literature can be expressed within this framework.  
The solution to the cosmological constant problem going by the name of ``sequester''~\cite{padilla,pad}, 
in its local formulation~\cite{pad1}, is an example of 
one of these theories. The ``fluxes'' defined in~\cite{pad1}  are nothing
but the age of the Universe according to two possible times, associated with deconstantized parameters. 
Indeed the stabilized {\it observed} cosmological constant in the sequester becomes the ratio
of two such ages. 
%This matter is currently under investigation. 
In this construction one takes the basis of constants:
\be
{\bm \alpha}=\left(\frac{1}{16\pi G_N},\rho_{0}=\frac{\Lambda}{8\pi G_N}\right)
\ee
(where $G_N$ is Newton's gravitational constant, we recall), i.e.
the Planck mass squared and the bare vacuum energy $\rho_0$, respectively. 
This leads to times' formulae:
\be
\partial_\mu {\bm T}^\mu_{\bm \alpha}=\sqrt{-g}\left(-R, 1\right),
\ee
(where $R$ is the Ricci scalar)
that is, unimodular time, and a version thereof weighted by the Ricci scalar, let's call it
Ricci time. This leads to two ages of the Universe (considering a past and future boundary defined in~\cite{pad}),
$\Delta T_{\alpha1}$ the Ricci time age, and $\Delta T_{\alpha 2}$ the volume time age. 
The space-time average of the Ricci scalar can be written as a ratio between these two ages:
\be
\langle R\rangle=-\frac{\Delta T_{ 1}}{\Delta T_{  2}}.
\ee
After some manipulations~\cite{pad1} it is then proved that the observed stabilised cosmological constant
is given precisely by:
\be
 \Lambda_{obs}=\frac{1}{4}\langle R\rangle =-\frac{\Delta T_1}{4 \Delta T_2}.
\ee
In a future publication we will investigate the connection between the results in this paper and the sequester. 
Other similar prescriptions targeting the gravitational coupling and the Planck constant were considered in~\cite{vikman,vikman1}.

\section{Reduction to minisuperspace}{\label{MSSred}}

It is straightforward to prove that these theories reduce to~\cite{Myletter} in minisuperspace.
We take for base action the Einstein-Cartan action reduced to homogeneity and isotropy (e.g.~\cite{CSHHV,MZ})
\begin{equation}\label{Sg}
S_0=6\kappa V_{c}\int 
dt\bigg(\dot{b}a^{2}-Na \left[- (b^{2}+k 
%-c^2 -\frac{\Lambda }{3}%
%a^{2}\bigg)
)+ \sum_i\frac{m_i}{a^{1+3w_i}}\right]\bigg),
%dt\bigg(a^{2}\dot{b}+Na (b^{2}+k +...
%-c^2 -\frac{\Lambda }{3}%
%a^{2}\bigg)
%)\bigg).
\end{equation}
where the last term describes a set of generic perfect fluids with equation of state $w_i$
(to be initially examined one at a time). 
Here $\kappa=1/(16\pi G_N)$, $k=0,\pm1$ is the normalized spatial curvature, 
%and $\Lambda$ is the cosmological constant. Here 
$a$ is the expansion factor
(the only metric variable), and the connection variable $b$ is
%and $c$ are components of the connection, respectively the
the off-shell version of the Hubble parameter (since $b= \dot a$ on-shell, if there is no torsion).
%, and the parity-violating
%Cartan spiral staircase~\cite{spiral,MZ}. 
The Lagrange multiplier $N$ is the lapse function and $V_c=\int d^3 x$ is the comoving volume of the region 
under study (which could be the whole manifold, should this be compact).

The summation  term  can accommodate a large number of models, but their details will not be relevant here. 
% and we will at first consider one at a time.  
For the cosmological constant  we have 
$m_i=\Lambda/3$ and  $w_i=-1$. For dust and radiation we have 
$w_i= 0,1/3$, and we can set $m_i=C_i 8\pi G_{N0}/3$, where $C_i$ are conserved quantities, such as those defined in~\cite{Gielen}, and 
the gravitational coupling is kept fixed $G_N=G_{N0}$. 
But we can also set  $m_i=C_i 8\pi G_N/3$, and deconstantize the gravitational coupling, $G_N$, instead.
%\footnote{ And dintinguish between this gravitational coupling, and the fixed  As opposed to the $G_{N0}$ controlling the non-commutation properties of  geometrical operators.  We may need to define $\kappa$ with a fixed fiducial $G_{N0}$~\cite{Myletter}  or not~\cite{bruno1},
%depending on whether the Planck mass itself becomes a target (as in~\cite{pad1}).}. 
None of these details (leading to alternative
theories) will be relevant to the solutions to be found here.

Hence, the Poisson bracket associated with the base action is:
\begin{equation}\label{PB1}
\{b,a^{2}\}=\frac{1}{6\kappa V_{c}},
\end{equation}%
%and since the constraints are always first class, this implies commutation relations:
leading to commutator: 
\begin{equation} \label{com1}
\left[ \hat{b},\hat{a^{2}}\right] =i\frac{l_{P}^{2}}{3V_{c}}=i\plk,
\end{equation}%
where $l_{P}=\sqrt{8\pi G_{N}\hbar }$ is the reduced Planck length, implying  an effective
``Planck's constant''\fn{Throughout the paper we will use the shorthand $ \plk$ or not depending on convenience, and comparison with previous literature. Its interpretation as the actual Planck constant~\cite{BarrowPlanck} will not be relevant here.}:
\be
\plk=\frac{l_P^2}{3V_c}.
\ee
In the $b$ representation (\ref{com1}) can be implemented by:
\be\label{a2inb}
\hat a^2=-i\frac{l_P^2}{3V_c}\frac{\partial}{\partial b}=i\plk \frac{\partial}{\partial b}.
\ee
We now focus on the case of a single fluid with equation of state $w$ (or an epoch where a fluid dominates all the others).
From $S_0$, a Hamiltonian $H_0$ can be derived:
\bea\label{totalHfluid}
H_0&=& 6\kappa V_c Na\left(-(b^{2}+k) +\frac{m}
{a^{1+3w}}\right),
\eea
which leads to the standard WdW equation:
\be\label{wdw}
H_0\psi_s(b,m)=0
\ee
with suitable ordering (namely that implied by (\ref{totalHfluid})).  A possible way to solve (\ref{wdw}) is to solve instead
$
{\cal H}_0\psi_s(b,m)=0
$
with 
\be\label{calHam}
    {\cal H}_0= h_\alpha(b)a^2- \alpha=0,
\ee
where:
\bea
h_\alpha(b)&=&(b^2+k)^{\frac{2}{1+3w}}\label{hm}\\
\alpha&=& m^{\frac{2}{1+3w}},\label{alpham}
\eea
that is, to solve:
\be\label{WdW}
\left[-i\plk h_\alpha(b)\frac{\partial}{\partial b}-\alpha\right]\psi_s(b;\alpha)=0.
\ee
This is a standard way to get a solution in the connection representation in the case of $\Lambda$ (as we review in the next Section), and generalizes trivially for radiation, and for other fluids (Section~\ref{Genfluids}) (in multi-fluid cases some subtleties may arise; see Section~\ref{crossover} and~\cite{b-bounce}). It leads to the (real) Chern-Simons state and its adaptations. Setting:
\be\label{Xalphadef}
X_\alpha (b)=\int \frac{db}{h_\alpha(b)}
\ee
the WdW Eq.~(\ref{WdW}) becomes a plane-wave equation in $X_\alpha$:
\be
\left(-i\frac{l_P^2}{3V_c} \frac{\partial}{\partial X_\alpha}-
 \alpha
\right)\psi_s =0,
\ee
with solution:
\be\label{planew0}
\psi_s(b;\alpha)={\cal N} \exp{\left[i\frac{3V_c}{l_P^2} \alpha X_\alpha(b) \right]}.
\ee
Notice that we use the subscript $\alpha$ to index $X_\alpha$ not because it is a function of $\alpha$, but because the function $X(b)$ depends on the $\alpha$ targeted.

Having established the base theory (no extension has been applied yet), we now subject the theory to prescription (\ref{Utrick}), targeting $\alpha$ (suitably normalized by $6\kappa$ for convenience). Hence, in minisuperspace:
\be\label{Sext1}
S_0\rightarrow S=S_0+6\kappa V_c\int dt \dot\alpha T_\alpha.
\ee
This addition does not change the Hamiltonian constraint, but it does introduces a new pair of canonical variables:
\be\label{PBa1}
\{\alpha,T_\alpha\}=\frac{1}{6\kappa V_c}.
\ee
so that :
\be\label{PB1}
\left[\alpha , T_\alpha\right]=i\plk.
%\frac{l_{P}^{2}}{3V_{c}},
\ee
It has the virtue of converting the WdW equation (\ref{WdW}) into a Schrodinger equation
\be\label{WDWSchro1}
\left[-i\plk h_\alpha(b)\frac{\partial}{\partial b}- i \plk \frac{\partial }{\partial T_\alpha}\right]\psi(b,T_\alpha) =0,
\ee
with the wave function now depending on time $T_\alpha$. 
Its monochromatic solutions are:
\be\label{monoch}
\psi(b,T_\alpha) = \psi_s(b;\alpha ) \exp{\left[-\frac{i}{\plk}\alpha T_\alpha \right]}
\ee
with the ``spatial'' $\psi_s$ satisfying the original (\ref{WdW}). The full monochromatic solutions are therefore plane-waves in $X_\alpha$ 
moving at fixed speed (set to 1 by the $6\kappa$ normalization of $\alpha$):
\be\label{planew0}
\psi (b, T_\alpha)= {\cal N}  \exp{\left[i\frac{3V_c }{l_P^2} \alpha (X_\alpha(b) - T_\alpha)  \right]},
\ee
with the choice 
\be
{\cal  N}=\frac{1}{\sqrt{2\pi\plk}}
\ee
to be justified later. 
Notice that $X_\alpha(b)$ is like a linearizing variable in DSR: the speed of propagation is variable if measured in terms
of the more physically available $b$, rather than $X_\alpha$.
The most general solution is a superposition of monochromatic solutions:
\be
\psi (b,T_\alpha)=\int \frac{d\alpha}{\sqrt{2\pi \plk}} {\cal A}(\alpha) \exp{\left[\frac{i }{\plk}\alpha (X_\alpha(b) - T_\alpha)  \right]}.
\ee

Note that the free Schrodinger equation (\ref{WDWSchro1}), is in fact a 
wave equation accepting only retarded waves:
\be
%-i\frac{l_P^2}{3V_c}
\left(\frac{\partial}{\partial X_\alpha}+
 \frac{\partial}{\partial T_\alpha}
\right)\psi =0.
\ee
%i.e. (notice this BIG difference - there are NOT two modes!!!!!).
Its associated conserved current is: 
%SHALL WE WORK OUT THE CONSERVED CURRENT FOR THIS? It appears to be:
\be\label{current}
j^0=j^1=|\psi|^2
\ee
i.e. all the waves are outgoing (or retarded time) solutions.  The general solution takes the form:
\be\label{FTXalpha}
\psi(b)=F(T_\alpha -X_\alpha),
\ee
where $F$ can be any function. At once a definition of probability is suggested, but we defer the matter to Section~\ref{norma-inner}.

%******* [MOVE THIS OR CUR IT]

%We have deferred to *** applying the deconstantization procedure of *** because...

%****************************************************

\section{Pure Lambda and a reinterpretation of Chern-Simons time}\label{Lambda}
We first illustrate these principles with the cosmological constant $\Lambda$, showing that the implications
are a twist on both unimodular gravity~\cite{unimod1,unimod} (specifically the  time variable defined 
in~\cite{unimod}), and the concept of Chern-Simons time~\cite{Chopin}. Indeed our proposal leads to a hybrid between
these works, with a significant reinterpretation of Chern-Simons ``time''. 

We start by reviewing some standard results.
For a pure $\Lambda$ we have in minisuperspace (ignoring torsion~\cite{GB,MZ}):
% [explain the ``matter'' action to be added]:
\bea\label{hamLamb}
H&=& 6\kappa V_c Na \left(-(b^{2}+k)+ \frac{\Lambda}{3}a^2 \right).
\eea
%We can also write this Hamiltonian in the form (\ref{totalHfluid}), setting $m=2\kappa\Lambda$.
A direct solution to the quantum Hamiltonian constraint in the $b$ representation:
% is given by:
\be
\left[-(b^{2}+k)-i \frac{\Lambda l_P^2}{9V_c}\frac{\partial}{\partial b}\right]\psi=0
\ee
is given by the (real) Chern-Simons state~\cite{CS,jackiw,kodama} reduced to minisuperspace~\cite{GB,MZ,CSHHV}:
\be\label{CS}
\psi_{CS}={\cal N}\exp{\left[i \frac{9V_c}{\Lambda l_P^2}
\left(\frac{b^3}{3}+bk\right).
\right]},
\ee
As is well known, this 
is a pure phase, which is the product of a ``frequency'' proportional to  $1/\Lambda$, and  
the  ``Chern-Simons time'' as proposed by Smolin and Soo~\cite{Chopin}.

Something similar can be obtained from our construction.  We can put the Hamiltonian constraint associated
with (\ref{hamLamb}) in the form:
\be
{\cal H}_0= \frac{1}{b^2+k}a^2-\frac{3}{\Lambda}=0.
\ee
%This is not a lapse function redefinition: it is a different function of the phase space variables that vanishes when 
%the original Hamiltonian vanishes. [SEE LATER; YES IT IS, ON-SHELL!!! ALSO CHANGE THE FACTOR IN CAL H]
% [exceptions? work them out]. 
%The point is that the Hamiltonian constraint now has 
By doing this the Hamiltonian constraint acquires the form (\ref{calHam}) with:
\bea
h_\alpha(b)&=&\frac{1}{b^2+k}\\
\alpha&=&\phi=\frac{3}{\Lambda}.
\eea
A Schrodinger equation (\ref{WDWSchro1}) follows, with a time variable $T_\phi$ identified with 
$p_\phi$, normalized such that:
\be
\left[\phi , T_\phi \right]=i\frac{l_{P}^{2}}{3V_{c}}\equiv i\plk 
\ee
(cf. (\ref{Sext1}) and (\ref{PB1})). Its monochromatic solutions are:
\bea
\psi &=& \psi_s(b;\phi ) \exp{\left[-i\frac{3 V_c }{l_P^2} \phi T_\phi \right]}
%\nn\\ &=&  \psi_s(b;\phi )   \exp{\left[-i\frac{9 V_c }{l_P^2\Lambda } T_\phi  \right]}.
\eea
with 
%$T_\phi\equiv p_\phi$ and with 
the ``spatial'' factor of the wave-function satisfying:
\be
\left[-i\frac{l_P^2}{3V_c} h_\alpha(b) \frac{\partial}{\partial b} - \frac{3}{\Lambda} \right]\psi_s=0.
\ee
This is the point of bringing the Hamiltonian to form (\ref{calHam}) and choosing the ordering we chose. 
As in (\ref{planew0}), the $\psi_s$ are plane waves:
\be\label{psi0CS}
\psi_s(b;\phi)={\cal N}  \exp{\left[i\frac{3V_c }{l_P^2} \phi X_\phi (b) \right]}.
\ee
in ``spatial'' variable:
\be\label{XCS}
X_\phi(b)=\int \frac{db}{h_\alpha(b)}=\frac{b^3}{3}+kb.
\ee 
This is nothing but the Chern-Simons functional (in minisuperspace) and (\ref{psi0CS}) is the Chern-Simons state (\ref{CS}). 

However, our interpretation of ``Chern-Simons time'' is different from that of Smolin and Soo. 
The full monochromatic solution is:
\be\label{monochL}
\psi (b,T_\phi)=  {\cal N} \exp{\left[i\frac{3V_c }{l_P^2 } \phi(T_\phi -X_\phi (b)) \right]}
\ee
with $T_\phi\equiv p_\phi$. Hence the (unitary) time evolution happens in terms of a time which is {\it not} the 
Chern-Simons functional, but the momentum conjugate to $1/\Lambda$ (up to a conventional proportionality constant). 
Here $X_i=\Im(Y_{CS})$ is not a time, but a spatial variable. Time, instead, is the conjugate of $\alpha=3/\Lambda$. 
The waves, however, move at constant speed (set to one by the conventional proportionality factors)
in terms of $X_\phi$ and $T_\phi$. This is true of the phase speed and also of the group speed if we construct
wave packets, as we shall do in the next Section. Hence the spatial $X_\phi$ and the time $T_\phi$ can be 
loosely confused  if $\psi$ is peaked.  Its peak moves along the outgoing ``light-ray'' $T_\phi=X_\phi$, and hence
confusing the two may be harmless for some purposes. 

%[cf. unimodular; bring in Misner time somewhere] 

%if we use variable $X_\phi $ 
%in terms of a spatial linearizing variable (cf. DSR) which is indeed what Smolin a Soo called the Chern-Simons time. 
%Their Chern-Simons time is in fact a spatial variable, equal to time {\it on the light cone} defined by these waves. 
%Their speed of propagation in terms of the connection, $b$, however, is necessarily variable. 

\section{More general states} \label{genstats}
By demoting $\Lambda$ to a circumstantial constant we gain more than a time variable in the quantum theory:
we enlarge the space of solutions. Instead of being restricted to (\ref{monochL}) we can now admit the 
most general superposition of these monochromatic plane waves:
\be\label{gensolL}
\psi(b)=\int \frac{d\phi}{\sqrt{2\pi\plk}} {\cal A}(\phi) \exp{\left[\frac{i}{\plk } \phi (X_\phi (b)-T_\phi ) \right]},
\ee
with the probability for the cosmological constant given from:
\be\label{Probphi}
P(\phi)=|{\cal A}(\phi)|^2
\ee
with measure $d\phi$ (fast-forward to the end of this Section for more details; also see~\cite{GBMSS} for alternatives).

Everything we state in this Section about general states for $\Lambda$, parameterized by $\phi$, works for any other $\alpha$, with suitable
modifications (i.e. $\phi\rightarrow \alpha$ in all relevant quantities).

\subsection{Extreme cases}\label{extremesols}
Two limiting cases are of interest. At one extreme we may have a completely undetermined $\phi$:
\be
{\cal A}(\phi)= \epsilon 
\ee
%[fix the measure $d\phi$ if needed.] 
leading to:
\be\label{rayXT}
\psi=\sqrt{2\pi\plk}\epsilon \delta(T_\phi-X_\phi).
\ee
%(which is actually required in~\cite{GB,MZ}, and expresses the conformal constraint).
This is the conformal constrain present in the parity-even branch of the  quasi-topological theories of~\cite{GB,MZ,GBMSS}, %based on the Gauss-Bonnet invariant,
where $\Lambda$ is allowed to vary by virtue of multiplying a Gauss-Bonnet topological term. In such a theories $1/\Lambda$ has a conjugate momentum which is forced to equal the Chern-Simons functional by a primary constraint. Here we see that this constraint is interpreted as a 
``light-ray'' in minisuperspace:
of the many waves generally acceptable only a delta function ray is possible in this theory. 
Time $T_\phi$ is fully fixed by the position $X_\phi$ along this ray. Whereas in standard Relativity
any state (\ref{gensolL}) is a solution, in these quasi-topological theories one is forced to have an infinitely 
sharp clock, with total delocalization in ``constant'' $\phi$.

At the opposite extreme we may completely fix $\phi$:
\be
{\cal A} (\phi)=\delta(\phi- \phi_0)
\ee
leading to:
\be
\psi ={\cal N}  \exp{\left[-i\frac{3V_c\phi_0 }{l_P^2} (T_\phi -X_\phi(b)) \right]}.
\ee
This is the Chern-Simons state in the usual Einstein-Cartan theory, where Lambda is fully fixed. It implies a 
uniform distribution in $X_\phi$. The crests of this infinite plane wave still move at the speed light,
but its ``location'' does not, because it is not localized. Hence time effectively disappears, since the wave
function is fully smeared in $X$ and $T$. 
This is an example of a more general fact: {\it infinitely sharp constants are failed clocks}. 
They imply complete delocalization in time. This clarifies and reinterprets the discussion on 
a time-like tower of turtles in~\cite{towerturtles}. Obviously these states are not strictly speaking normalizable.

%Connect this with the Universe that lost the time [paper with Lee]\fn{Would this be a good place to  point out the issue of: the CC problem vs. a function of a fine tuned variable might not be fine tuned. What is the measure? Here it is 1/CC, so the question is not why it is so small, but why it is so large?}.
\subsection{Coherent squeezed states}\label{Seccoherent}
In between these two extremes we can build coherent/squeezed states centred at $\phi_0$:
\be\label{Amp}
{\cal A}(\phi)=\sqrt{{\bf N}(\phi_{0},\sigma_\phi)}=
\frac{\exp{\left[-\frac{(\phi-\phi_0)^2}{4\sigma_\phi^2 }\right]}}
{(2\pi \sigma_\phi ^2)^{1/4}},
\ee
(where $\bf N$ denotes a normal distribution). 
Evaluating the integral we get:
\bea\label{coherent}
\psi(b,T)&=& {\cal N}'
\exp\left[\frac{-\sigma_\phi^2(X_\phi-T_\phi)^2+i \phi_0(X_\phi-T_\phi)}{\plk^2}\right]
%[i\frac{3V_c }{l_P^2} \phi (X_\phi (b)-T_\phi )
\nn\\
&=&{\cal N}'\psi(b,T_\phi;\phi_0)\exp\left[-\frac{\sigma_\phi^2(X_\phi-T_\phi)^2}{\plk^2}\right].
\eea
The last expression relates the infinite norm Chern-Simons state for $\phi_0$ to the
finite norm wave packet built around a fixed $\Lambda$.  We see that it is dressed by a Gaussian, which regularizes it.
%Indeed the associated  probability density is:
%\be
%P(b)=|\phi(b)|^2={\cal N}^2\sqrt{8\pi\sigma_\phi_c^2}\exp{\left[-\frac{18 b^2\sigma_\phi_c^2}{\Lambda^2\plk^2}\right]}.
%\ee
This is just a Gaussian distribution in $X-T$, with variance:
\be
\sigma_T^2=\sigma_X^2=\frac{ \plk^2}{4\sigma_\phi^2}.
\ee
A Heisenberg uncertainty principle can therefore be established, 
with:
$\sigma_X=\sigma_T$ and 
\be
\sigma_T\sigma_\phi\ge  \frac{ l_P^2}{ 6V_c }.
\ee
For a coherent squeezed state this inequality is saturated.

Note that here (as in~\cite{zlos}) there is an ambiguity in defining zero squeezing. Requiring 
%\be\label{cohsigma}
$\sigma_T^2=\sigma_\phi^2 =  l_P^2/( 6V_c )$ would be dimensionally wrong. 
%\ee
This is nothing but the the ambiguity in defining coherent non-squeezed states for free particles~\cite{freecoh}\fn{ 
Our deconstantized constants can be seen as the momentum of a free abstract particle, moving with uniform speed in
an abstract ``space'' which is our time variable.}.
It is well known that, unlike for a harmonic oscillator or EM radiation, coherent states for a free a particle lack a natural scale with which to define
dimensionless quadratures~\cite{knight,freecoh} and the squeezing parameter. One may ignore this
and simply look at squeezed-coherent states as a general class of solutions, or else introduce a scale in the problem (as was done
in~\cite{zlos}).

\subsection{Solitons}
Note also that we do not need to use monochromatic waves to build states. As already 
explained, any function of the form:
\be\label{genF}
\psi=F(X_\phi-T_\phi)
\ee 
would work, as we saw in the discussion leading to (\ref{FTXalpha}).
Namely $F$ can be just a Gaussian, without the plane wave factor, as for (\ref{coherent}):
\bea\label{soliton}
\psi(b,T)&=& {\cal N}'' \exp\left[\frac{(X_\phi-T_\phi)^2}{4\sigma^2 }\right]
\eea
Such ``solitons'' could be interesting. We stress that unlike coherent states, such solitons
are not well localized in $\phi$: for that we need the internal beats of a plane-wave, for which 
this $F$ would be the envelope.

\subsection{Normalizability and measure}\label{norma-inner}
All of these solutions are normalizable with the ``naive'' inner product.
No longer do we need to blame the trivial inner product
for the non-normalizability of the monochromatic solutions. Of course monochromatic solutions are strictly speaking non-normalizable by themselves; their superpositions, on the other hand, are normalizable in the standard sense.

Specifically, mimicking the procedure in~\cite{vilenkin} for the simpler current (\ref{current}),  we can infer the inner product:
\be\label{innX}
\langle\psi_1|\psi_2  \rangle=\int dX_\phi  \psi_1^\star(b,T_\phi)\psi_2(b,T_\phi)
\ee
with unitarity:
\be
\frac{\partial} {\partial T_\phi}\langle\psi_1|\psi_2  \rangle=0
\ee
enforced by current conservation:
\be
\frac{\partial}{\partial T_\phi}\langle\psi_1|\psi_2  \rangle=\int dX_\phi \frac{\partial}{\partial X_\phi}(\psi_1^\star(b,T_\phi)\psi_2(b,T_\phi))=0.
\ee
(This vanishes only with suitable boundary conditions, with subtleties like the ones highlight in~\cite{Gielen}, e.g. singularities, etc.)
We can also swap $X_\phi$ and $T_\phi$ in this definition:
\be\label{innT}
\langle\psi_1|\psi_2  \rangle=\int dT_\phi \psi_1^\star(b,T_\phi)\psi_2(b,T_\phi)
\ee
with the two definitions equivalent and amounting  to:
\be\label{innalpha}
\langle\psi_1|\psi_2  \rangle=\int d\phi  {\cal A}_1^\star (\phi) {\cal A}_2(\phi) ,
\ee
in view of (\ref{gensolL}). The normalizability condition $|\langle\psi |\psi   \rangle|=1$ therefore supports (\ref{Probphi}) 
identifying the probability of $\phi$.

We stress that this argument is valid for each dominating fluid $i$, adopting its associated $\alpha$, $X_\alpha$ and $T_\alpha$. 
The argument for unitarity is far more complicated in the transition regions for multi-fluids, or for the minority clock, i.e. 
clocks corresponding to sub-dominant components, as we shall see later in this paper. 

%[TBD ADD A DISCUSSION OF THE RANGE OF X: VERY IMPORTANT]

%and this is currently being investigated\footnote{Preliminary investigations suggest that (\ref{innalpha}) generalizes to these situations, although this is no longer  equivalent to (\ref{innX}).}.

%In follow up work we will in fact use as the natural inner produ

%COPY RELEVANT BIT FROM QUANTU TORSION, CHECK THE FACTS OF 3 THERE AND ADAPT HERE 

%  [Unitary evolution - comment]

%The GB theory required $A(\phi)=1/(2\pi)$. The fixed Lambda theory. The Lambda with uncertaintity. 

%\subsection{Contrast with unimodular}

\section{Radiation and general perfect fluids}\label{Genfluids}
Our approach can be used to find the equivalent of the Chern-Simons state
for Universes filled with a single fluid with an equations of state more general than $w=-1$.

\subsection{The example of radiation}
A radiation dominated Universe ($w=1/3$) is a particularly simple case. 
Then the Hamiltonian is:
\bea\label{hamrad}
H&=&6\kappa V_c   \frac{N} { a} \left(-(b^{2}+k)a^2 + m \right)
\eea
already in the form (\ref{calHam}, up to a redefinition of the lapse function:
\be
\tilde N= 6\kappa \frac{N}{a} V_c
\ee
even off-shell. 
Time, therefore, is the conjugate momentum to $m$ and this can be identified with conformal time
[more on this later].  The monochromatic solutions to the time-dependent Sch equations are:
\be
\psi = \psi_s(b;m) \exp{\left[-\frac{i}{\plk}mp_m\right]}
\ee
with:
%[shall we just present the solutions here or even earlier?]
%Secondly, with the ordering implied in (\ref{hamrad}) and using (\ref{a2inb}) we can derive the relation:
\be
\left[i\plk (b^2 +k)\frac{\partial}{\partial b}+m\right]\psi_s=0.
\ee
so that we have plane waves in terms of:
%This can be solved directly, with solutions: ***.
%But we can also introduce the variable $X$ such that:
\bea
X_r(b)=\int \frac{db}{b^2+k}=&\frac {1}{\sqrt{k}}\arctan [\frac{b}{\sqrt{k}}]&\qquad{\rm if}\; k>0\nn\\
=&-\frac{1}{b}& \qquad{\rm if}\; k=0\nn\\
=&-\frac {1}{\sqrt{|k|}}{\rm argtanh} [\frac{b}{\sqrt{|k|}}]&\qquad{\rm if}\; k<0\nn
\eea
to be seen as the equivalent of the Chern-Simons functional for a radiation dominated Universe.
The plane wave solutions at a generic time therefore are:
\be
\psi(b,T_r;m)={\cal N} \exp{\left[\frac{i}{\plk}m (X_r - T_r) \right]}.
\ee
(with $T_r=p_m$). These solutions will form the basis for the solution of the singularity problem proposed in~\cite{gielensing}. 
%and the general solution:
%\be
%\psi=\int dm A(m) \exp{\left[-\frac{i}{\plk}m (T-X) \right]}.
%\ee
%[NORMALIZATION????]

%We thus have two extreme cases here, Lambda and radiation. 
%A solution to the Lambda problem (and possibly flatness, etc) could be the result
%of conflating these 2 solutions. Is this an S-matrix problem? Or a diffraction-like problem
%(since the speed of propagation in $b$ space changes). 

\subsection{One exception: Milne or curvature domination}\label{milne-exception}
The general solution (\ref{hm})  breaks down for $w=-1/3$, an equation of state degenerate with spatial curvature $k$ (or $kc^2$, to put it suggestively). 
Backtracking to (\ref{totalHfluid}) we see that the problem is that we lose the spatial differential operator contained in $a^2$.
The spatial solution then becomes $\psi_s=\delta(b^2-m)$, where $m$ can include, or indeed be just  $-kc^2$. 
The monochromatic solution is:
\be
\psi(b,T_m ;m)={\cal N} \delta (b^2-m)e^{-\frac{i}{\plk} m T_m}
\ee
with $T_m=p_m$ as usual. 
%This might be one case where it is better to use the $a^2$ representation, for which ETC.
Hence in this case there is no time evolution, since for any superposition we have:
\bea
\psi(b,T_m)&=&\int dm {\cal A}(m) \delta (b^2-m)e^{-\frac{i}{\plk} m T_m}\nn\\
&=&{\cal A}(b^2) e^{-\frac{i}{\plk} b^2 T_m}
\eea
so that $|\psi|^2=|{\cal A}(b^2)|^2$. The reason why this happens will be made clear in the next Section.

\section{The classical limit}\label{classlimit}
Given that ``time evolution'' is the most obvious feature of classical cosmology, it is obvious that any quantum cosmology scheme 
lacking a ``time'' will have trouble connecting with the classical world.  Reciprocally, the discovery of a quantum time should be used 
in the first instance to make sure that the classical limit is sound, before exploring possible quantum departures/corrections. 

In this Section we first present 
the format in which the classical results are best presented so that they can be recovered by appropriately (semi)-classical
wave-functions, within our scheme. We then prove that coherent states reproduce the classical limit. 

\subsection{The classical ``time-formula''}\label{sec-timef}
We first find a classical  expression for our physical times $T_\alpha$ as a function of the non-physical coordinate time $t$ associated 
with lapse $N$. 
%\subsection{To be shifted to a general discussion on time}
%[SHIFTed from  EARLIER, TOGETHER WITH THE BIT JUST ABOVE]
%This can be seen immediately for radiation and Lambda. 
From the second Hamilton equation (using (\ref{totalHfluid}),  (\ref{PBa1}) and (\ref{alpham})) we can derive  the ``time-formula'':
\bea\label{timeformula}
\dot T_\alpha=\dot p_\alpha =\{p_\alpha,H\}&=&-\frac{1+3w}{2}Na^{-3w} m^\frac{3w-1}{1+3w}\nn\\
&=&-\frac{1+3w}{2}Na^{-3w} \alpha ^\frac{3w-1}{2}.
%\dot T_\alpha= -\frac{\partial H}{\partial \alpha},
\eea
Note that we have used the original Hamiltonian, and not ${\cal H}_0$, to work out the relation between 
$T_\alpha$ and time. 
We can now set $N=1$ to derive the relation between $T_\alpha$ and cosmological proper time $t$:
\be
\frac{dT_\alpha}{dt}=-\frac{1+3w}{2} a^{-3w} m^\frac{3w-1}{1+3w}
\ee
or else set 
\be
N=N_\alpha=-\frac{2}{1+3w} a^{3w} m^\frac{1-3w}{1+3w}
\ee
to ensure we are using a time coordinate coincident with the physical time $T$. 

Within this scheme (but note~\cite{bruno1}), we highlight the following facts:
\begin{itemize}
\item Radiation is unique in that its time does not depend on $m$, so when this goes to zero
its time is still well defined. 
\item Specifically, radiation time is minus conformal time, $T_r=-\eta$, since:
\be
\dot T_r=-N/a.
\ee
This is in agreement with~\cite{Gielen}. 
\item Dust time is proportional to minus proper cosmological time, with:
\be
T_m=-\frac{t}{2m}.
\ee
\item Our Lambda time is proportional to unimodular time~\cite{unimod1}: 
\be
\dot T_\phi=N\frac{ a^3}{\phi^2}=N\frac{\Lambda^2}{9}a^3.
\ee
A canonical transformation relates the two: this is responsible for
linearizing the dispersion relations. Unimodular time is related to Misner's volume time~\cite{misner}: indeed it can be seen as a 4D version, where time is measured by the 4-volume to the past of an observer.  
\item The only degenerate case in (\ref{timeformula}) is $w=-1/3$, but this case is exceptional, as already discussed in Section~\ref{milne-exception}.
In this case we should not transform from $m$ to $\alpha$, so that $\dot T_m=-Na$. 
\end{itemize}

The sign in the time-formula (\ref{timeformula}) is important and we note that it changes from $w>-1/3$ to $w<-1/3$.
This is a key feature of our formalism and we will comment further on this below.

% and related to the time formulation of the horizon problem REPETITION? CUT?
%It's true that we could switch the sign of the momentum, but if we stick to one convention it would seem that 
% $\dot T$ changes sign if $w>-1/3$ or $w<-1/3$. THE ARROW OF TIME MUST BE COMMENTED UPON.

%[Funny that for matter ($w=0$), $T$ is proportional to cosmological time: $T=-t/(2m)$. ]

%Implications for the LeeMag uncertainty principle? 

\subsection{The classical trajectory: a connection space picture}
All classical descriptions are equivalent, so we may select whichever makes
better contact with our quantum theory. In our case we pick a description which is unusual in that:
\begin{itemize}
\item Instead of  the expansion factor $a$, we take for dependent variable the minisuperspace connection variable $b$.
Recall that when torsion is zero, on-shell this is the comoving inverse Hubble length $\dot a=b$. Quantum mechanically $b$
is an independent and complementary variable to the metric (or rather, the densitized inverse triad $a^2$). 
%$b$, in line with connection space  primacy.
\item Instead of using some coordinate time $t$ as independent variable, we use the physical time(s)  $T_\alpha$.
These are classically (on-shell) a function of $t$, as just calculated in (\ref{timeformula}). Fundamentally, and quantum mechanically,
there can be many $T$, but in the classical limit they all become functions of $t$ (so that there is only one time classically, but
there are several quantum mechanical times). 
\end{itemize}
Hence, the classical description we are aiming for takes the 
form $b=b(T)$, possibly in the parametric form:
\bea
b&=&b(t)\\
T_\alpha&=&T(t),
\eea
rather than the textbook $a=a(t)$.

Then, we can show that the classical trajectory for a single fluid system is given by:
\be\label{dotXdotT}
\dot X_\alpha=\dot T_\alpha.
\ee
Indeed the full content of the classical equations can be obtained from the first Friedman equation
(equivalent to the Hamiltonian constraint $H=0$):
\be\label{F1}
b^2+k= \frac{m}
{a^{1+3w}}
\ee
which should be assumed throughout, 
as well as the two dynamical Hamilton equations:
\bea
\dot a&=&\{a,H\}=Nb\\
\dot b&=&\{b,H\}=-\frac{1+3w}{2}N a\frac{m}{a^{3(1+w)}}\label{F2}
\eea
(where we have used (\ref{F1}) in the second equation). Together, these two 
dynamical equations imply the Raychaudhuri (second Friedman) equation (for $N=1$):
\be\label{F2a}
\ddot a=-\frac{1+3w}{2}a\frac{m}{a^{3(1+w)}}.
\ee
It is easy to see that (\ref{dotXdotT}) implies:
\be
\frac{\dot b}{h_\alpha(b)}=-\frac{1+3w}{2} a^{-3w} m^\frac{3w-1}{1+3w}
\ee
which upon some manipulations reproduces the dynamical equation (\ref{F2}) (assuming 
the constraint (\ref{F1}) throughout). 

%\bea
%\dot b=\ddot a &=&-\frac{1+3w}{2}a\frac{m}{a^{3(1+w)}}\nn\\
%&=&-\frac{1}{2}\frac{1}{6\kappa}a(\rho+3p).
%\eea
%[UPDATE NOTATION]

Using this unconventional description (i.e. (\ref{dotXdotT}))  may take some getting used to, even though it is 
classically equivalent to the $a=a(t)$ description. Points of note include:
\begin{itemize}
\item Expanding and contracting Universes correspond to $b>0$ and $b<0$, with $b=0$ representing a
static Universe (and its vicinity the loitering model). 
\item Hence, the ekpyrotic~\cite{ekp}, or any such similar ``bouncing'' models, will see $b$ go through zero. Tunnelling between 
branches with different signs may also be possible. 
\item 
%The time-reversible $|b|$ (or  increases in ``time'', regardless of the arrow of time
For a given matter content, $b$ can either only increase or only decrease in parameter time $t$; the first if $w<-1/3$, the second
if $w>-1/3$. For $w=-1/3$ (or for the Milne Universe, for example) $b$ does not change (this starts shedding light one the
anomaly found in Section~\ref{milne-exception}).
\item Hence, the equivalent in of a ``bounce'' in $b$ space is a Universe undergoing a transition from 
decelerated to accelerated expansion, such as we have seemingly undergone recently. At the end of inflation 
the reverse happens, the equivalent of a ``turn-around'' in $b$ space. 
\end{itemize}

The fact that in this picture we have recently emerged from a $b$-bounce must have quantum mechanical implications:
quantum reflection always leaves its traces. The matter will be studied in more detail in~\cite{b-bounce}.

\subsection{Parenthesis on the ``arrow of time''}
In view of what we said,  the issue of the arrow of ``time'' merits a parenthesis. 
In the metric representation flipping the time arrow inter-converts expanding (increasing $a$) and contracting
 (decreasing $a$) Universes. 
%[but $a^2$ always increases]. 
There are always two branches of solutions, as required by time-reversal symmetry. 

In the  $b=b(T)$ description, face value, there is only one solution, for which $b$ must increase with $T_\alpha$:
\be\label{dbdT}
\frac{db}{dT_\alpha}>0,
\ee
% (for a single fluid)\fn{This can also be understood
%form the fact that F2 is invariant under $t\rightarrow -t$.}. 
as  implied by (\ref{dotXdotT}).  This is reflected in the quantum mechanical  solutions (cf. (\ref{FTXalpha})): there can only be outgoing waves. 
Thus, there is a sense in which there is only one arrow of time in the connection representation and using the times $T_\alpha$.
A Feynman absorber is not needed to set the physical arrow of time.

This fact is actually an expression of the horizon (and ultimately also the flatness) problem, as well as its standard 
solution, as we now show. Let us first assume 
expanding Universes ($b>0$). Then, the horizon problem is that for $w>-1/3$ the comoving Hubble length $1/b$ increases in time,
whereas its solution follows from that it decreases  if  $w< -1/3$. In our description this is expressed by
(\ref{dbdT})  and the fact that, using $t$ as the auxiliary arbitrary time arrow, the sign
in the time-formula (\ref{timeformula}) changes from $w>-1/3$ to $w<-1/3$.  Converting this into $b(T_\alpha)$ we then 
get that (\ref{dbdT}) is a statement of the horizon problem for $w>-1/3$ and its solution for $w<-1/3$. 
%The outgoing-only wave is then an expression of the horizon problem in that $b$ must always increase with $T_\alpha$ for all $w$.
%The fact that the arrows of the $T_\alpha$ are classically reversed ensures that $b$ must decrease for  $w>-1/3$ and increase for  $w>-1/3$.

The actual arrow of $t$ does not matter, because it cancels in its double effect on $b=\dot a$ and on $\dot b$ (note the invariance
of the Raychaudhuri equation (\ref{F2a}) under time reversal).  Solutions to the horizon problem 
based on a contracting phase (e.g. the ekpyrotic scenario~\cite{ekp}) can be understood in our description from 
(\ref{dbdT}) being independent of  the sign of $b$; however the statement of the problem and its solution involves
$|b|$, not $b$.  So the criteria for problem and solution are reversed for models in a contracting phase ($b<0$), 
and this is still expressed by (\ref{dbdT}).

%[CUT? ADD? SHIFT TBD]

%Quantum mechanically the times are all independent, as well as their arrows. Classically, they become all a function of each other, and so their {\it relative} arrow must be fixed. Radiation and matter times must always have the opposite arrow to Lambda time. 

%Using $t$ as the auxiliary arbitrary time arrow, we see that, as we pointed out already, the signin the time-formula (\ref{timeformula}) is important (by which we mean the relative sign for the two cases), and changes  from $w>-1/3$ to $w<-1/3$.  The outgoing-only wave is then an expression of the horizon problem in that $b$ must always increase with $T_\alpha$ for all $w$. The fact that the arrows of the $T_\alpha$ are classically reversed ensures that $b$ must decrease for  $w>-1/3$ and increase for  $w>-1/3$.

. 

%\subsubsection{Multifluids and multi-times}

\subsection{Coherent states and the classical limit}
Having a quantum time variable and a larger space of solutions (as described in Sec.~\ref{genstats})
are the two reasons why contact with the classical limit is possible. 
Monochromatic waves, such as those solving the fixed constant theory, imply a uniform distribution (in $X(b)$), hardly a prediction, 
but they are also not immediately physical.  One needs {\it both} a time variable {\it and} 
the ability to superpose plane waves into normalized peaked distributions to recover something minimally physical.

In fact having a peak is not enough. For example, since $\dot X_\alpha=\dot T_\alpha$ represents the classical trajectory
(as just discussed), one might think that the light-ray, $\psi\propto\delta(X_\alpha-T_\alpha)$, described in Section~\ref{extremesols}
would be perfectly classical.
But such a state would have a totally undefined $\alpha$, and so the $T_\alpha=T_\alpha(t)$ part of the argument could not be true
(note that $\alpha$ generally appears in the RHS of (\ref{timeformula})). 
%Is the uniform distribution in $b$ with fixed $\alpha$ be the smoothing in time of the trajectory?

The semi-classical limit is only recovered for the coherent states $\psi(b,T_\alpha)$ described in Section~\ref{Seccoherent}.
For these, the  second Hamilton equation (\ref{timeformula})
is true not only on average (an expression of Ehrenfest's theorem), but with minimal and balanced 
uncertainties in the complementary $\alpha$ and $T_\alpha$ appearing on the two sides of (\ref{timeformula}).
Both sides of the argument implying that $\dot X_\alpha=\dot T_\alpha$ represents the classical trajectory can now
be reproduced, and so we have a truly semi-classical state.

We can also understand the result in Section~\ref{milne-exception}. We do not have propagating waves in this case. 
However, the classical equation of motion is $\dot b=0$, i.e. the Universe is static in $b$, as already explained. 
Any coherent state in $m$ therefore reproduces this result (as well as the time formula in terms of $m$).

\section{Multitime}\label{multitime}
Naturally, we end up with the usual problem in Quantum Gravity: either there is no time, or, if we succeed in defining one, we are left 
with a multitude of choices. 
How do we deal with multiple times and multiple fluids, even in situations where there are epochs where
one fluid dominates? 
The proposal in~\cite{Myletter} was to accept this multitude of times, with the adjustment of clocks across different ``time zones'' to be seen as a physical  feature of our world. For the rest of this paper we will examine further  how the handover between clocks can be made seamless for some states. 
%But there are also alternative states, which might be  of great predictive relevance, considering that the Universe is currently busy passing on the baton from a matter (or $G_N$) clock to a $\Lambda$ clock, a matter to be studied elsewhere~\cite{b-bounce,matLambda}.

%But now we have an embarrassment of riches.
Let $\bm{\alpha}$ be a vector with dimension $D$ representing the whole set of 
relevant constants and  $\bm{T}$ their conjugates. The $D$ components of $\bm{T}$
are a priori independent variables, so we have a plethora of times instead of a single one\footnote{We used ``a priori'' here because ``a posteriori'',
i.e. on-shell or semiclassically, all the $T$ become a function of each other, as we will see later, so that this is not to be confused with classical theories with extra time
dimensions.}. Hence the ``Schrodinger'' equation is a PDE in multiple times 
%running concurrently,
obtained from taking the Hamiltonian following from (\ref{Sg})
%\begin{equation}
%H= Na \left[- (b^{2}+k 
%-c^2 -\frac{\Lambda }{3}%
%a^{2}\bigg)
%)+ \sum_i\frac{m_i}{a^{1+3w_i}}\right],
%\end{equation}
and applying the replacement:
\be\label{genSchrod}
H\left[b,a^2;\bm \alpha \rightarrow  i \frac{l_P^2}{3V_c} \frac{\partial }{\partial \bm T}\right]\psi=0
\ee
(in whatever representation, $a^2$, or $b$ as chosen here). 
Its general solutions are:
\be\label{gensol}
\psi(b,\bm T)=\int d\bm{\alpha} {\cal A}(\bm \alpha) \exp{\left[-i\frac{3V_c }{l_P^2}\bm \alpha \bm T  \right]}\psi_s(b;\bm \alpha),
\ee
where $\psi_s(b;\bm \alpha)$ solves the WDW equation with constant $\bm\alpha$. 
We fix:
\be
|\psi_s|^2=|{\cal N}_D|^2=\frac{1}{(2\pi\plk)^{D}}
\ee
to streamline the algebra.
% of the equivalent of Section~\ref{norma-inner} for multi-time. 

As outlined in~\cite{Myletter}, 
if the Hamiltonian divides phase space into regions dominated by a single constant (or a single fluid), the readjustment of quantum clocks 
across such regions is seamless if we assume coherent states in all $\alpha_i$ {\it and}  factorization:
\be\label{factcoh}
{\cal A}(\bm \alpha) =\prod_i\sqrt {{\bf N}(\alpha_{0i},\sigma_i)}.
\ee
Then, as explained in~\cite{Myletter},
%and for a coherent 
%${\cal A}(\bm \alpha)$ in all $\bm \alpha$ (so that the boundaries between such regions are clearcut), 
the $\psi_s(b;\bm \alpha)$ is a piecewise plane wave in the $X_{i}(b)$ associated with each dominant $\alpha_i$. 
Inserting into (\ref{gensol}), each of these pieces gets grouped into a factor with the  phase associated with the corresponding $T_{i}$ 
(hence, the approximate single-time Schrodinger equation),
producing a wave-packet describing the correct classical limit (as described before).  

In the next Sections we will make this argument more explicit, whilst addressing details such as 
what happens in the transition regions between single fluid domination, or what happens to the minority component clocks 
in each phase where one  fluid dominates. In order to address these questions in detail, and properly deal with multi-fluid situations, we need 
to first introduce a new tool for generating solutions in minisuperspace.

%If ${\cal A}(\bm \alpha)$ factorizes,  all the other times factorize too, and stop describing the $X$ evolution.  We end up with a classical  limit in terms of different $X$ and $T$ in each region, but classically they are all equivalent. 

\section{Minisuperspace as a dispersive medium}\label{MSSDSR}
The arguments in Section~\ref{MSSred} for linearizing  variables $\alpha$, $X_\alpha$ and $T_\alpha$ are straightforward to apply only 
when one fluid dominates  the Universe. The real world, however, is more complicated. 
%Unfortunately the real world does not oblige. 
For example, at the crossover between 2 epochs dominated by different fluids, we will find an $X$ variable which is a 
function not only of $b$ but also of $\alpha$ (as can be seen from the prescription leading to (\ref{Xalphadef}), and will be found explicitly in 
the  next Section). 
In such cases it is more fruitful to revert to the original variable $b$ (instead of any function $X(b)$) and regard minisuperspace as a dispersive 
medium. From this point of view the variables $\alpha$, $X_\alpha$ and $T_\alpha$, when they exist, are the ``linearizing variables'' of the 
dispersive medium, to use the terminology of~\cite{DSR}. They can and should be used where they exist, but
more generally we should face the dispersive nature of minisuperspace head on.

In the general case we can still define times $\bm T$ for the various  $\bm \alpha$, impose the monochromatic ansatz
(\ref{gensol}), and find the spatial solutions $\psi_s$, which in general 
will not be plane waves in any $X(b)$ variable independent of $\alpha$. The monochromatic solutions can still be superposed
into peaked wave-packets, as in (\ref{gensol}). However it is important to realize that, as with any other dispersive medium, 
the envelope of such packets moves with a group speed that should not be confused with the phase speed.

Specifically, writing:
\be
\psi_s(b,{\bm \alpha})={\cal N}_D\exp{\left[i\frac{3V_c }{l_P^2} P(b,{\bm \alpha})\right]}
\ee
we identify dispersion relations:
\be
\bm \alpha \cdot \bm T-P(b,\bm \alpha)=0.
\ee
Assuming that the 
amplitude ${\cal A}({\bm \alpha})$ is factorizable and sufficiently peaked around ${\bm \alpha}_0$,
we can expand:
\be
P(b,\bm \alpha)=P(b;\bm \alpha_0)+\sum_i\frac{\partial P}{\partial \alpha_i}\biggr\rvert_{\bm \alpha_0}(\alpha_i-\alpha_{i0})+...
\ee
to find that the wave function factorizes as:
\be\label{approxwavepack}
\psi\approx 
{\cal N}_D
%\exp{\left(
e^{ i\frac{3V_c }{l_P^2} (P(b;\bm\alpha_0)-\bm \alpha_0\cdot \bm T)}
%\right)}
\prod_i \psi_i(b,T_i).
\ee
The first factor is the monochromatic (generally non-plane) wave centered on $\bm\alpha_0$. The other factors
describe envelopes of the form:
\be
\psi_i(b,T_i)=\int d\alpha_i{\cal A}(\alpha_i) e^ {-i\frac{3V_c }{l_P^2} (\alpha_i-\alpha_{i0})(T_i- \frac{\partial P}{\partial \alpha_i}
%\biggr\rvert_{\bm \alpha_0}
)}
\ee
which therefore move according to:
\be
T_i=\frac{\partial P(b)}{\partial \alpha_i}\biggr\rvert_{\bm \alpha_0}.
\ee
We can also dot this equation, to find the group speed on 
$\{b,T_i\}$ space:
\be\label{cg}
c_g=\frac{d b}{d T}\biggr\rvert_{peak}=\frac{\dot b}{\dot T}\biggr\rvert_{peak}=\frac{1}{\frac{\partial ^2P}{\partial \alpha\partial b}}.
\ee
The motion of these envelopes (and so of the peak of the distribution) should agree with the classical equations of motion.
We will show in the rest of this paper that indeed it does so, for coherent states, in a number of non-trivial situations
(such as for mixtures of fluids during transition periods when none of them dominates, or for the sub-dominant clock).

This obviously generalizes the construction for single fluids, for which a variable $X(b)$ can be found such that 
$P=\alpha X(b)$ for some $\alpha$. Then, with a a suitable choice of  $\alpha$ (and canonical $T_\alpha$)
we can always make the first term in the dispersion relations $\alpha T_\alpha$ (for example, in the case of Lambda by 
$\Lambda\rightarrow \phi=3/\Lambda$, $T_\Lambda\rightarrow T_\phi =- T_\Lambda/\phi^2$). 
They are ``linearizing'' variables because  $c_{lin}={\dot X}/{\dot T}=1$.

%since when they exist $P=\alpha X(b)$ implying 
%\be
%\frac{\dot b}{\dot T}\biggr\rvert_{peak}=\frac{1}{X'(b)}
%\ee
%or $\dot X=\dot T$, as required. 

%ADD THIS? IT BREAKS THE FLOW AND MIGHT NOT BE RELEVANT HERE

%[ADD? As for $\alpha$ and $T_\alpha$ this is only a convenient canonical transformation. Convenient but meaningless for example in the case where a single constant is used at different epochs, see $G_N$ paper.  Do we need to generalize this?]

%[INSERT COMMENTS HERE ON CONSTRAINTS BETWEEN CLOCKS?] 

\section{Dealing with crossover regions}\label{crossover}
As it happens we are sitting right on a bounce in $b$. How do we deal with such transitions?
In this Section we show that the correct semi-classical limit is still obtained  {\it assuming 
the wave function remains sharply peaked}. What actually happens to the wave function is left to future work~\cite{b-bounce}.
We also investigate the fate of the minority clock (i.e. the radiation and Lambda clocks in the Lambda and radiation epochs) 
once the handover of clocks is completed.
We will use as a working model a mixture of radiation and Lambda, because the algebra is clearer, but generalizations 
to the more relevant case of dust and Lambda behave in the same way. 
%[can we do this directly with them? or  at least adapt at the end?]

\subsection{Mono-chromatic solutions}
Our working model has classical Hamiltonian:
\be
H= N a\left (-(b^2+k) +  \frac{a^2}{\phi}+ \frac{m}{a^2}\right)
\ee
spanning  a two-dimensional constant space with 
\be
{\bm \alpha}=\left(\phi\equiv \frac{3}{\Lambda},m\right).
\ee
Its multi-time ``Schroedinger'' equation (\ref{genSchrod}) has solutions of the form (\ref{gensol}). One way to find the 
spatial $\psi_s$ is to  put $H=0$ in the form (\ref{calHam}) with $\alpha=\phi$,
%(we could do this with $\alpha=m$ too), 
aware that $h_\alpha(b,\alpha)$ may then be a function of $\alpha$ too. To this end we  
solve the quadratic in $a^2$ equivalent to $H=0$ to find:
\be
a^2=\frac{g\pm\sqrt {g^2-4m/\phi}}{{2}/{\phi}}
\ee
with $g(b)=b^2+k$. 
Since $a^2$ must be real (although not necessarily positive)
we have:
\be
g^2\ge g_0^2=\frac{4m}{\phi}=\frac{4}{3}\Lambda m.
\ee
The plus branch contains Lambda domination when $g^2\gg g_0^2$; the 
minus branch contains radiation domination, also with $g^2\gg g_0^2$. The transition happens
when $g^2\approx g_0^2$ (with $g^2>g_0^2$). 
%[ADD A PLOT?]
%(The two branches are nothing but an expression of the horizon problem and its usual 
%solution in inflationary cosmology.)
Thus,
%(with $g>0$, i.e. an expanding universe), 
we have a ``bounce'' in $b$ space at $g=g_0$, i.e. 
a transition from decelerated expansion (decreasing $b$) to accelerated expansion  (increasing $b$). 
The Hamiltonian constraint  is therefore equivalent to two constraints of the required form:
\be
{\cal H}_\pm= h_\pm (b;\phi,m)a^2-\phi=0
\ee
with the important novelty that $h$ (and so $H_0$) is ``energy''-dependent (dependent on the 
conjugate of time; i.e. the constants):
\be
h_\pm=\frac{2}{g\pm\sqrt{{g^2- 4 m/\phi}}}.
\ee
%regardless of which  time we choose.
This is of course irrelevant for the $\psi_s$, which is given by:
\be
\psi_{s\pm}(b;\phi,m)={\cal N}  \exp{\left[i\frac{3V_c }{l_P^2} \phi X_\pm  (b;\phi,m) \right]}.
\ee
with:
\be\label{XLm}
X_\pm(b;\phi, m)=\int db\frac{1}{2}\left(g\pm\sqrt{{g^2- 4 m/\phi}}\right).
\ee
We see that for $g^2\gg m/\phi $ the $+/-$ branches have:
\bea
X_+(b;\phi, m)&\approx&X_\phi=\frac{b^3}{3} + kb\\
X_-(b;\phi, m)&\approx&\frac{ m}{\phi}X_r
\eea
leading to the correct limits:
\bea
\psi_{s+}(b;\phi,m)&\approx &{\cal N}  \exp{\left[i\frac{3V_c }{l_P^2} \phi X _\phi  (b) \right]}\\
\psi_{s-}(b;\phi,m)&\approx &{\cal N}  \exp{\left[i\frac{3V_c }{l_P^2} m X_r  (b) \right]}.
\eea
This illustrates with a concrete example the comments made just after Eq.~(\ref{factcoh}): the $\psi_s(b;\bm \alpha)$ is a 
piece-wise plane wave in the relevant $\alpha$ and $X_\alpha$ in each region of single fluid domination. To leading order it might seem that
if ${\cal A}(\bm \alpha)$ factorizes, then all the other times factorize, too, and stop describing the $b$ evolution since they became $b$-independent
phases.  However this is not the case, as we now show by considering the next to leading order in the expansion.

\subsection{What happens to the minority clock(s)?}
Before addressing the handover region itself, we first examine in more detail what happens to 
the ``minority'' clock once the handover is finished. 
%This issue is particularly relevant for quantum states where one of the constants is perfectly sharp, so that its conjugate is a failed clock, in which case we have to rely on the minority clock. 
%To leading order it might seem that the minority clock becomes just a phase
Expanding  (\ref{XLm}) to the next order we find:
\bea
X_+(b)&\approx&X_\phi -\frac{m}{\phi}X_r+... \\
X_-(b)&\approx&\frac{ m}{\phi}X_r+\frac{m^2}{\phi^2}\int \frac{db}{g^3}+ ... .
\eea
%It is then instructive to see what happens to the radiation and Lambda clocks deep in the Lambda and radiation epoch, respectively.
%as this illustrates different situations. 

\subsubsection{The radiation clock in the Lambda epoch}\label{min1}
Including the next order term in $X_+$ we find that 
deep in the Lambda epoch the monochromatic wave function is:
\be
\psi(b,{\bm T}; {\bm \alpha})=
{\cal N}  \exp{\left[-i\frac{3V_c }{l_P^2} (\phi (T_\phi-X _\phi )  
+ m (T_r+X_r ))\right]}\nn.
\ee
%(where the + is not a typo). 
Inserting into (\ref{gensol}) we find for any factorizable amplitude:
\be\label{jointF}
\psi(b,{\bm T})= F_1(X_\phi -T_\phi)F_2(X_r+T_r).
\ee
In particular we could choose factorizable Gaussian amplitudes for $\phi$ and $m$ leading to 
coherent $F_1$ and $F_2$ (of the form (\ref{coherent})). As we will see this is more the exception than the rule.

We see that {\it both} factors reproduce the classical equations of motion.  These amount  to the first and second Friedmann equations:
\bea
b^2+k&=&\frac{a^2}{\phi}+\frac{m}{a^2}\label{F1phim}\\
\dot b&=&\frac{a}{\phi}-\frac{m}{a^3}.\label{F2phim}
\eea
In addition, the times formulae (replicated quantum mechanically by coherent factorizable states, just as before) are:
\bea
\dot T_\phi&=&\frac{a^3}{\phi^2}\\
\dot T_r&=& -\frac{1}{a}.
\eea
Evaluating:
\bea
\dot X_\phi&=&\dot b (b^2+k)\\
\dot X_r &=&\frac{\dot b} {b^2+k}
\eea
we can then recover the mono-fluid equations of motion in the appropriate epochs with 
$\dot X_\alpha\approx \dot T_\alpha$, for $\alpha=\phi,m$.  But some more algebra also reveals that 
deep in the Lambda era we can write the classical trajectory as:
\be\label{minradLamb}
\dot X_r\approx -\dot T_r
\ee
and this is equivalent  to  $\dot X_\phi\approx \dot T_\phi$.  Hence, the peak of both factors in (\ref{jointF})
describes the classical trajectory.

This sheds light on what happens to our quantum ``multi-time'' in semi-classical situations, 
given that classically only one time can exist. 
Quantum mechanically the two $T_i$ are independent variables and fundamentally remain
so, even for semi-classical states. There is never a constraint between the different $T_i$. What happens
is that for peaked states the peak of the joint distribution maps out a trajectory of $b$ in 2D space
$\bm T$ (in this case $X_\phi(b) =T_\phi$ {\it and} $X_r(b) =- T_r$). This implies a constraint between the two times
{\it at the peak of the joint  distribution}, so classically only one time exists. The quantum fluctuations or these different times, on the other hand,
would remain independent.

%[SHIFT COMMENT ON REVERSE ARROW HERE?]

%Note the classical equations of motion are equivalent to  both $\dot X_\phi=\dot T_\phi$ {\it and}
%$\dot X_r=-\dot T_r$. 
%[SPELL OUT MORE]

%[HOW TO ARTICULATE THIS?]

%The fact that the peak in the join $\phi$ requires a peak in both factors contains the constraint 
%that classically there is only one time, that is

\subsubsection{The Lambda clock in the radiation epoch}\label{minoLambda}
Deep in the radiation epoch we have instead
\bea
&&\psi(b,{\bm T}; {\bm \alpha})=\label{minoL}\\
&&{\cal N}  \exp{\left[-\frac{i}{\plk} \left(m (T_r-X _r )  
+ \phi \left(T_\phi- \frac{m^2}{\phi^2}\int \frac{db}{g^3} \right)\right)\right]}\nn
\eea
with the novelty that the factor associated with the subdominant component (Lambda) now depends on $m$ as well. 
As a result, the wave packets never factorizes into separate radiation and Lambda factors, even if the  
amplitudes ${\cal A}(\bm \alpha)$ do. In addition the minority factor no longer is a plane wave in the original $X(b)$ and 
$\alpha=\phi$. Hence, even if the original amplitudes were a diagonal Gaussian, the wave functions will be very distorted. 
The simple arguments for unitarity for single fluids also break down for a minority clock in this situation. 
This will be discussed further in the next Section. 

Nonetheless, we can show that for a peaked second factor, the motion of the peak still reproduces the correct classical limit
(the first one obviously does).
Using (\ref{cg}) and 
\be
c_g^{-1}= \frac{\partial ^2}{\partial \phi \partial b} \frac{m^2}{\phi }\int \frac{db}{g^3}=-\frac{m^2}{\phi^2 g^3}
\ee
we find that for the peak:
\be
c_g= \frac{\dot b}{\dot T_\phi}\biggr\rvert_{peak}
\ee
implies
\be
\dot b=- \frac{m}{a^3}
\ee
which is nothing but approximately (\ref{F2phim}) in the radiation epoch.

Notice that within the same approximations used in Section~\ref{MSSDSR}) to derive (\ref{cg}), the wave-function effectively 
factorizes as: \be
\psi(b,{\bm T}; {\bm \alpha})
\approx \psi_1(T_r-X_r)\psi_2(b,T_\phi;m_0)
\ee
Hence to this order, the arguments at the end of Section~\ref{min1} in support of a single classical time still apply.
%and in fact we do assume that the wave-function remains approximately factorizable and coherent. Whether this is a good approximation remains to be seen~\cite{b-bounce}. 

\subsection{Semiclassical limit in the transition region}
%Obviously while the crossover is taking place the expressions are more co
We can also show that {\it if} the distribution remains peaked~\cite{b-bounce},
%[something we will examine in detail elsewhere]
then the peak follows the classical trajectory even during the $b$-bounce, for both branches $\pm$. In this case the $P$ function 
defined in Section~\ref{MSSDSR} is given by:
\bea
P_\pm(b,m,\phi)&=&\phi X_\pm (b; m/\phi)\nn\\
&=&\phi \int db\frac{1}{2}\left(g\pm\sqrt{{g^2- 4 m/\phi}}\right).
\eea
Within the approximations of Section~\ref{MSSDSR} (see (\ref{approxwavepack}) in particular) the wave function must be approximately 
given by the monochromatic solution times the product of two envelopes $ \psi_1(b,T_\phi)\psi_2(b,T_r)$. 
The latter move with group speeds:
\bea
c_{g1}&=& \frac{\dot b}{\dot T_\phi}\biggr\rvert_{peak}=\frac{1}{\frac{\partial ^2 P}{\partial \phi \partial b} }\\
c_{g2}&=& \frac{\dot b}{\dot T_r}\biggr\rvert_{peak}=\frac{1}{\frac{\partial ^2 P}{\partial m  \partial b} }.
\eea
It is now a matter of algebra to shown that in both branches ($\pm$) these are equivalent to the classical equation
of motion (\ref{F2phim})  (with (\ref{F1phim})  assumed throughout). 

Indeed, for the Lambda wave packet factor we have: 
\be
{\frac{\partial ^2 P}{\partial \phi \partial b} }=\frac{1}{h}\pm\frac{m}{\phi}\frac{1}{\sqrt{g^2-g^2_0}}.
\ee
Using:
\be
\pm \sqrt{g^2-g^2_0}=\frac{2a^2}{\phi}-g=\frac{a^2}{\phi}-\frac{m}{a^2}
\ee
and $h=\phi/a^2$ we have:
\be
{\frac{\partial ^2 P}{\partial \phi \partial b} }=\frac{a^4/\phi^2}{\frac{a^2}{\phi}-\frac{m}{a^2}}
\ee
implying that the peak moves along:
\be
\dot b=\frac{\dot T_\phi}{a^4/\phi^2}\left(\frac{a^2}{\phi}-\frac{m}{a^2}\right)=\frac{a}{\phi}-\frac{m}{a^3}
\ee
i.e. (\ref{F2phim}),  as required. 

Likewise, for the $m$ wave packet factor we have:
\bea
{\frac{\partial ^2 P}{\partial m \partial b} }&=&\mp \frac{1}{\sqrt{g^2-g^2_0}}\nn\\
&=&-\frac{1}{\frac{a^2}{\phi}-\frac{m}{a^2}}
\eea
leading to:
\be
\dot b=- \dot T_r{\left(\frac{a^2}{\phi}-\frac{m}{a^2}\right)}
\ee
or (\ref{F2phim}) . 

Hence the correct classical limit is always obtained, assuming the wave functions remain peaked. 
Whether this is a good approximation remains to be seen~\cite{b-bounce}. In addition there are other issues
regarding the semi-classical limit, as we now explain.

\section{Why swap clocks?}\label{why-swap}

In this Section we explain better why ``a clock [should be] crafted with what is at hand'', as proposed in~\cite{Myletter}.
This is not just common sense: It affects the semi-classical limit. As we have just seen in detail, the probability peak's motion has 
the correct classical limit (assuming Ehrenfest's theorem)  even for the minority clock, but this hides the fact that typically the state will
not be coherent in such a set up, and so departures from the semi-classical regime are expected.

The Lambda clock in the radiation epoch is a good illustration of this. As we saw in Section~\ref{minoLambda}
(cf. Eq.~\ref{minoL}),  to leading order (in the saddle approximation of Section~\ref{MSSDSR}), the Lambda factor in the radiation epoch 
changes its dependence on $b$ from $X_\phi$ to:
\be
X= \int \frac{db}{g^3}
\ee
and its $\alpha$ from $\phi$ to:
\be
\alpha=\frac{m_0}{\phi^2}.
\ee 
In Section~\ref{minoLambda} we studied in detail the peak of the wave function, but the semi-classical limit requires also the 
arguments in Section~\ref{classlimit} for the correct representation of the time-formula, and so we need more than a peaked distribution: 
we should have a coherent state in $m_0/\phi^2$. But if we chose a Gaussian amplitude in $\phi$, then this will not be Gaussian in $m_0/\phi^2$,
quite the contrary: strong distortions are expected. This is representative of what usually happens to minority clocks. Indeed, the radiation clock in the Lambda epoch (see Section~\ref{min1}) is  the exception to this rule. It is a rare case where a coherent majority clock remains coherent in the subdominant phase.

%[CUT THIS PARAGRAPH?]
%The example of $G_N$: a single clock through different epochs. All good in terms of variables. Just a canonical transformation.  The problem is the quantum theory. The packets can only be coherent and saturate the H relations in one epoch. It must necessarily  depart from it somewhere in that case. 

We can also add the issue of unitarity and inner product to the discussion. We can define a conserved inner product as in Section~\ref{norma-inner},  but it only leads to a simple conserved current and re-expression in terms of a measure in $b$ in the dominant epoch. 
As we saw, in mono-fluid situations (and so using the dominant clock in a multi-fluid situation) there is a range of options for setting up the inner product and conserved current. These are all ultimately equivalent: we can either use Eq.~(\ref{innX}), leading to general expression:
\be
d\mu(b)=dX_\alpha=\frac{db}{(b^2+k)^{\frac{2}{1+3w}}}, 
\ee
or we can trade $X_\alpha$ for $T_\alpha$ leading to (\ref{innT}), or we can use (\ref{innalpha}) (with $\phi$ replaced by the applicable $\alpha$). 
However, only
\be\label{innalpha}
\langle\psi_1|\psi_2  \rangle=\int d\alpha  {\cal A}^\star (\alpha_1) {\cal A}(\alpha_2) .
\ee
generalizes to multi-fluid situations.
% with a particularly simple result if we use only the dominant clock.
Bearing in mind that the general solution now is:
\be\label{gensol-last}
\psi(b,T)=\int d{\alpha} {\cal A}( \alpha) \exp{\left[-i\frac{3V_c }{l_P^2} \alpha  T  \right]}\psi_s(b; \alpha),
\ee
where $\psi_s(b; \alpha)$ solves the WDW equation with constant $\alpha$, it is obvious that (\ref{innalpha}) reduces to 
(\ref{innX}) and  (\ref{innT}) for single fluids. For multi-fluids we recover  (\ref{innT}) iff $\psi_s(b; \alpha)$ is a pure phase
(so in cases where there is no $b$ bounce), but not (\ref{innX}). For multi-times this is still more complicated. But
whatever the case, the definition  (\ref{innalpha}) remains a general time-independent definition for the inner product. 
However, it is only when the dominant clock and its $X$ are used that it leads to a simple inner product in terms of $b$
(see~\cite{b-bounce} for further discussion).

%THIS IS ESSENTIAL, BUT IT MAY BE ARGUED THAT IFA. VARIABLE LOSES ITS CLASSICAL MEANING AS TIME VARIABLE, UNITARITY WITH RESPECT TO IT IS NOT IMPORTANT. 

\section{Conclusions}
In summary, in this paper we proposed 
%we started off by proposing a number of aspirational principles for defining clocks and rods in quantum cosmology,
an amplification of the standard theory allowing the constants of nature to be non-constant off-shell. 
Classically, the ``constants'' remain constant in the equations of motion, and so nothing changes, but
quantum mechanically it all changes. Each constant generates a space of quantum 
states composed of superpositions of waves indexed by the value of the constant (these can be seen as monochromatic partial waves). 
The waves propagate in a fundamentally dispersive medium where the ``space'' is the connection, the ``time'' is the momentum conjugate to the constant, and the ``energy'' and ``momentum'' are functions of the constant.  In some regions 
(or ``epochs'')
we can find simple linearizing variables ($X_\alpha$, $\alpha$ and $T_\alpha$), in terms of which the partial waves are plane waves moving at fixed speed, conventionally set to 1.  
%These implement  our aspirational principles. 
In such regions, for a given constant, our construction provides a good clock and rod, leading to a simple inner product and definition of unitarity, with coherent states providing the perfect definition of a semi-classical states. However, this construction is never global, and hence the 
need to change clocks at  different ``time zones'' in our Universe. 

Specifically, we showed that the dominant fluid always generates a good clock, but the ``minority clock'' is generally  problematic
in that the linearizing variables are not always available, and so the dispersive nature of the medium has to be faced. Even if approximate 
linearizing variables can be found, they will generally be different functions of the constant than in the dominant phase. 
If the wave function was coherent in the original variable, it will not be in the second. If the wave function is 
peaked, the peak follows the classical equations of motion; however the quantum nature of the system can never be erased.
Unitarity can be defined in a universal way, but it becomes cumbersome when written in terms of the minority clock. 
Changing clocks is therefore advisable. 

The interesting point remains that in transition regions we may expect anomalies, and so interesting phenomenology. 
To make matters more poignant, we happen to be loosely 
sitting on the fence separating matter and Lambda domination. Our clocks, therefore, have been likely non-ideal at one point
%non-aspirational for the past 
somewhere a few billion years ago. The fact that the transition from deceleration to acceleration is a quantum bounce,
with its inevitable ringing, only compounds the issue (see~\cite{b-bounce} for a preliminary investigation). This
was  followed up in~\cite{matLambda} in a more realistic setting and  in connection with the Hubble tension anomaly, 
providing the perfect arena for testing this idea observationally. 
%We should not regard this as an ontological tragedy but as an epistemological opportunity. s the Universe going quantum?

Obviously many questions remain. Our predictions in~\cite{b-bounce,matLambda} will be in the form of semiclassical corrections, but suppose
the Universe enters one of the non-semiclassical states also considered in this paper. 
%To this question we should attach another, more fundamental one:
How would we see a Universe going quantum? We are used to quantum systems as microscopical sub-systems living inside larger classical macroscopical systems; but this is just the opposite. What if our local classical world were encased in Universe which on the very largest scales is behaving quantum mechanically? How would we see it? We are currently investigating this matter within the context of 
the decoherent histories approach~\cite{Jonathan,Jonathan1}.

Even ignoring this question, other interesting problems remain. 
Clock swapping is essential for keeping the classical description, but why would the Universe choose to swap clocks?
A selection principle seems to be at play, and this may shed light on other fine tuning problems, such as the cosmological constant problem (e.g.~\cite{weinberg,padilla,pad,pad1,lomb,vikman,vikman1}). 
Also, what implications are there for the constants of nature if they are not allowed to be infinitely sharp in a classical world?
In our framework, an infinitely sharp constant implies total delocalization in the conjugate time. 
In this sense, a perfect constant is a failed clock\fn{One may also appeal to
Borges' quote, ``[Eternity can be defined]
as the simultaneous and lucid possession of all the instants of time".   
With this definition, an exact constant of Nature is equivalent to eternity {\it in one of the many possible times} (viz. the one
dual to the infinitely sharp constant).}. Would this have implications, either for our local physics, or for our description of the 
large-scale Universe (the two being essentially complementary)? Finally, one may ask what happens if more than one dominant clock is at play?
Preliminary work suggests that they would get in each other's way regarding classicality~\cite{bruno1}, but how problematic is this?
Is this, instead, another observational window of opportunity?

{\it Acknowledgments.} I would like to thank  Bruno Alexandre, Steffen Gielen, Jonathan Halliwell, Chris Isham and Tony Padilla 
for discussions and advice in relation to this paper. This work was supported by the STFC Consolidated Grant ST/T000791/1.

%\bibliographystyle{unsrtnat}
%\bibliography{references}

\begin{thebibliography}{99}


%EXCELLENT REVIEW!!!
\bibitem{Isham}
C.~J.~Isham,
``Canonical quantum gravity and the problem of time,''
NATO Sci. Ser. C \textbf{409}, 157-287 (1993).
%[arXiv:gr-qc/9210011 [gr-qc]]; 
\bibitem{Kuch} K. Kuchar,   ``Time and interpretations of quantum gravity," in 
Proceedings of
the 4th Canadian Conference on General Relativity and Relativistic Astrophysics',
World Scientic, Singapore, 1992.

\bibitem{marina}
M.~Cort\^es and L.~Smolin,
%``The Universe as a Process of Unique Events,''
Phys. Rev. D \textbf{90}, no.8, 084007 (2014); Phys. Rev. D \textbf{90}, no.4, 044035 (2014). 

\bibitem{rovelli}C.~Rovelli and M.~Smerlak,
%``Thermal time and the Tolman-Ehrenfest effect: temperature as the 'speed of time',''
Class. Quant. Grav. \textbf{28}, 075007 (2011).
%doi:10.1088/0264-9381/28/7/075007
%[arXiv:1005.2985 [gr-qc]].



\bibitem{Barrowtip} 
J.~D.~Barrow and F.~J.~Tipler,
``The Anthropic Cosmological Principle,'' Oxford University Press, 1988. 


\bibitem{Myletter}
J.~Magueijo,
%``Cosmological time and the constants of nature,''
Phys. Lett. B \textbf{820}, 136487 (2021)
doi:10.1016/j.physletb.2021.136487
[arXiv:2104.11529 [gr-qc]].


\bibitem{Duff}
M.~J.~Duff, L.~B.~Okun and G.~Veneziano,
%``Trialogue on the number of fundamental constants,''
JHEP \textbf{03}, 023 (2002)
doi:10.1088/1126-6708/2002/03/023
[arXiv:physics/0110060 [physics]].

\bibitem{unimod1}
W. G. Unruh, Phys. Rev. {\bf D40}, 1048 (1989). 
\bibitem{unimod} M. Henneaux and C. Teitelboim, Physics Letters {\bf B 222}, 195
(1989).

\bibitem{Bombelli}
L.~Bombelli, W.~E.~Couch and R.~J.~Torrence,
%``Time as space-time four volume and the Ashtekar variables,''
Phys. Rev. D \textbf{44}, 2589-2592 (1991)
doi:10.1103/PhysRevD.44.2589


\bibitem{UnimodLee2}
L.~Smolin,
%``Unimodular loop quantum gravity and the problems of time,''
Phys. Rev. D \textbf{84}, 044047 (2011)
doi:10.1103/PhysRevD.84.044047
[arXiv:1008.1759 [hep-th]].



\bibitem{misner}
C.~W.~Misner, Phys. Rev. {\bf 186}, 1328 (1969); Phys. Rev.1 {\bf 86}, 1319 (1969). 


%
%``Mixmaster universe,''
%Phys. Rev. Lett. \textbf{22}, 1071-1074 (1969).
%doi:10.1103/PhysRevLett.22.1071
%626 citations counted in INSPIRE as of 08 Mar 2021


\bibitem{Chopin}
L.~Smolin and C.~Soo,
%``The Chern-Simons invariant as the natural time variable for classical and quantum cosmology,''
Nucl. Phys. B \textbf{449}, 289-316 (1995). 
%doi:10.1016/0550-3213(95)00222-E
%[arXiv:gr-qc/9405015 [gr-qc]].


\bibitem{york} J. W. York, Phys. Rev. Lett. {\bf 26}, 1656 (1971); {\bf 28}, 1082 (1972).


\bibitem{bruno1}
%B. Alexandre and J. Magueijo, ``Semiclassical limit problems with concurrent use of several clocks in quantum cosmology'', 
B.~Alexandre and J.~Magueijo,
%``Semiclassical limit problems with concurrent use of several clocks in quantum cosmology,''
Phys. Rev. D \textbf{104}, no.12, 124069 (2021)
doi:10.1103/PhysRevD.104.124069
[arXiv:2110.10835 [gr-qc]].

%arXive: 2110.10835, to be published in Phys, Rev. D
% include here the discussion with G_n and fluids...
%REREF


%\bibitem{PlanckIsham} J. Magueijo and C. Isham, ``Quantizing the Planck constant: a self-referential construction'', in preparation.

\bibitem{CS}S. S. Chern and J. Simons, Ann. Math.  {\bf 99}, 48 
(1974);
G.~V.~Dunne,
``Aspects of Chern-Simons theory,''
[arXiv:hep-th/9902115 [hep-th]].


\bibitem{jackiw}
R.~Jackiw,
%``TOPOLOGICAL INVESTIGATIONS OF QUANTIZED GAUGE THEORIES,''
Conf. Proc. C \textbf{8306271}, 221-331 (1983)
MIT-CTP-1108.

%ADD THE JACKIW???
%\bibitem{jackiw}
\bibitem{kodama}
H.~Kodama,
%``Holomorphic Wave Function of the Universe,''
Phys. Rev. D \textbf{42}, 2548-2565 (1990);
J.~Magueijo,
%``Real Chern-Simons wave function,''
Phys. Rev. D \textbf{104}, no.2, 026002 (2021)
doi:10.1103/PhysRevD.104.026002
[arXiv:2012.05847 [gr-qc]].
%``The real Chern-Simons wave function,''
%[arXiv:2012.05847 [gr-qc]]. ADD PUBLISHED REF



\bibitem{GB}
S.~Alexander, M.~Cort\^es, A.~R.~Liddle, J.~Magueijo, R.~Sims and L.~Smolin,
%``Cosmology of minimal varying Lambda theories,''
Phys. Rev. D \textbf{100}, no.8, 083506 (2019);
Phys. Rev. D \textbf{100}, no.8, 083507 (2019). 

\bibitem{MZ}
J.~Magueijo and T.~Z\l{}o\'snik,
%``Parity violating Friedmann Universes,''
Phys. Rev. D \textbf{100}, no.8, 084036 (2019).



\bibitem{CSHHV}
J.~Magueijo,
%``Equivalence of the Chern-Simons state and the Hartle-Hawking and Vilenkin wave-functions,''
Phys. Rev. D \textbf{102}, no.4, 044034 (2020).
%doi:10.1103/PhysRevD.102.044034
%[arXiv:2005.03381 [gr-qc]].

\bibitem{GBMSS}
J.~Magueijo, T.~Zlosnik and S.~Speziale,
%``Quantum cosmology of a dynamical Lambda,''
Phys. Rev. D \textbf{102}, no.6, 064006 (2020)
doi:10.1103/PhysRevD.102.064006
[arXiv:2006.05766 [gr-qc]].



\bibitem{towerturtles}
%\bibitem{Magueijo:2018jzg}
J.~Magueijo and L.~Smolin,
%``A Universe that does not know the time,''
Universe \textbf{5}, 84 (2019)
doi:10.3390/universe5030084
[arXiv:1807.01520 [gr-qc]].
%14 citations counted in INSPIRE as of 13 May 2021

\bibitem{BarrowPlanck}
J.~D.~Barrow and J.~Magueijo,
%``A contextual Planck parameter and the classical limit in quantum cosmology,''
Found. Phys. \textbf{51}, no.1, 22 (2021)
doi:10.1007/s10701-021-00433-0
[arXiv:2006.16036 [gr-qc]].


\bibitem{Gielen} 
S.~Gielen and L.~Men\'endez-Pidal,
%``Singularity resolution depends on the clock,''
Class. Quant. Grav. \textbf{37}, no.20, 205018 (2020); S.~Gielen and L.~Men\'endez-Pidal,
%``Unitarity, clock dependence and quantum recollapse in quantum cosmology,''
[arXiv:2109.02660 [gr-qc]].
%doi:10.1088/1361-6382/abb14f
%[arXiv:2005.05357 [gr-qc]].

\bibitem{DSR}G.~Amelino-Camelia,
%``Relativity in space-times with short distance structure governed by an observer independent (Planckian) length scale,''
Int. J. Mod. Phys. D \textbf{11}, 35-60 (2002); 
%doi:10.1142/S0218271802001330
%[arXiv:gr-qc/0012051 [gr-qc]].
J.~Magueijo and L.~Smolin,
%``Lorentz invariance with an invariant energy scale,''
Phys. Rev. Lett. \textbf{88}, 190403 (2002).
%doi:10.1103/PhysRevLett.88.190403
%[arXiv:hep-th/0112090 [hep-th]].

\bibitem{b-bounce}
S.~Gielen and J.~Magueijo,
``Quantum analysis of the recent cosmological bounce in comoving Hubble length'', arXiv:2201.03596. 

\bibitem{gielensing}
%\bibitem{Gielen:2022ouk}
S.~Gielen and J.~Magueijo,
%``Quantum resolution of the cosmological singularity,''
[arXiv:2204.01771 [hep-th]].

\bibitem{matLambda}
%\bibitem{Alexandre:2022ijm}
B.~Alexandre and J.~Magueijo,
%``Possible quantum effects at the transition from cosmological deceleration to acceleration,''
[arXiv:2207.03854 [gr-qc]].
%B. Alexandre and J. Magueijo, ``Quantum cosmological effects at the transition from cosmological deceleration to acceleration'', to be submitted. 

%``Implications of a recent cosmological bounce'', 



\bibitem{vilenkin}
A. Vilenkin.
\newblock Quantum cosmology and the initial state of the universe.
\newblock \emph{Phys. Rev. D}, 37:\penalty0 888--897, Feb 1988.
\newblock \doi{10.1103/PhysRevD.37.888};
A. Vilenkin.
\newblock {Approaches to quantum cosmology}.
\newblock \emph{Phys. Rev.}, D50:\penalty0 2581--2594, 1994.
\newblock \doi{10.1103/PhysRevD.50.2581}.

\bibitem{ekp}
J.~Khoury, B.~A.~Ovrut, P.~J.~Steinhardt and N.~Turok,
%``The Ekpyrotic universe: Colliding branes and the origin of the hot big bang,''
Phys. Rev. D \textbf{64}, 123522 (2001)
doi:10.1103/PhysRevD.64.123522
[arXiv:hep-th/0103239 [hep-th]].


\bibitem{weinberg}
%\bibitem{Weinberg:1988cp}
S.~Weinberg,
%``The Cosmological Constant Problem,''
Rev. Mod. Phys. \textbf{61}, 1-23 (1989).
%doi:10.1103/RevModPhys.61.1

\bibitem{padilla}
%\bibitem{Padilla:2015aaa}
A.~Padilla,
``Lectures on the Cosmological Constant Problem,''
[arXiv:1502.05296 [hep-th]].


\bibitem{pad}
N.~Kaloper and A.~Padilla,
%``Sequestering the Standard Model Vacuum Energy,''
Phys. Rev. Lett. \textbf{112}, no.9, 091304 (2014).

\bibitem{pad1}
N.~Kaloper, A.~Padilla, D.~Stefanyszyn and G.~Zahariade,
%``Manifestly Local Theory of Vacuum Energy Sequestering,''
Phys. Rev. Lett. \textbf{116}, no.5, 051302 (2016)
doi:10.1103/PhysRevLett.116.051302
[arXiv:1505.01492 [hep-th]].

\bibitem{lomb}
L.~Lombriser,
%``On the cosmological constant problem,''
Phys. Lett. B \textbf{797}, 134804 (2019).


\bibitem{vikman}
P.~Jirou\v{s}ek, K.~Shimada, A.~Vikman and M.~Yamaguchi,
%``Losing the trace to find dynamical Newton or Planck constants,''
JCAP \textbf{04}, 028 (2021).
%doi:10.1088/1475-7516/2021/04/028
%[arXiv:2011.07055 [gr-qc]].

\bibitem{vikman1}
A.~Vikman,
%``Global Dynamics for Newton and Planck,''
[arXiv:2107.09601 [gr-qc]].

\bibitem{Jonathan} J. J. Halliwell, Probabilities in Quantum Cosmological Models: A Decoherent Histories
Analysis Using a Complex Potential," Phys. Rev. D80 (2009) 124032 (arXiv:0909.2597). 
\bibitem{Jonathan1}
 J. J. Halliwell,  Macroscopic Superpositions, Decoherent Histories and the
Emergence of Hydrodynamic Behaviour,  in  'Everett and his Critics', eds. S. W. Saunders et al. (Oxford University Press, 2009),
arxiv:0903.1802. 

\bibitem{knight}C. Gerry and P. Knight, Introduction to quantum optics, CUP, Cambridge 2004. 
\bibitem{freecoh}A. de la Torre and D. Goyeneche, 
``Coherent states for free particles'', 
[arXiv:1004.2620 [quant-ph]].

\bibitem{zlos}
J.~Magueijo and T.~Zlosnik,
%``Quantum torsion and a Hartle-Hawking beam,''
Phys. Rev. D \textbf{103}, no.10, 104008 (2021)
doi:10.1103/PhysRevD.103.104008
[arXiv:2012.07358 [gr-qc]].



\end{thebibliography}

\end{document}